\documentclass[10pt,conference]{IEEEtran}

\usepackage{amsmath,amssymb,amsfonts,mathtools}

\usepackage[T1]{fontenc}
\usepackage[utf8]{inputenc}
\usepackage{microtype}

\usepackage{libertinus}
\usepackage[libertine]{newtxmath}
\usepackage[normalem]{ulem}

\usepackage{graphicx}
\usepackage{booktabs}
\usepackage{multirow}
\usepackage{array}
\usepackage{tabularx}
\usepackage{siunitx}

\usepackage{algorithm}
\usepackage{algpseudocode}

\usepackage{cite}
\usepackage{xcolor}
\usepackage{enumitem}
\usepackage{balance}

\usepackage[caption=false,font=footnotesize]{subfig}

\usepackage[
    colorlinks=true,
    linkcolor=blue,
    citecolor=blue,
    urlcolor=blue
]{hyperref}

\newcommand{\ket}[1]{|#1\rangle}

\title{Towards the Characterization of Logical Errors in Distributed Lattice Surgery}

\author{
\IEEEauthorblockN{
Nitish Kumar Chandra\IEEEauthorrefmark{1}\IEEEauthorrefmark{2},
Reza Nejabati\IEEEauthorrefmark{2},
Eneet Kaur\IEEEauthorrefmark{2}
}

\IEEEauthorblockA{\IEEEauthorrefmark{1}
Department of Informatics \& Networked Systems, School of Computing \& Information\\
University of Pittsburgh, Pittsburgh, PA 15260, USA\\
Email: nkc16@pitt.edu
}

\IEEEauthorblockA{\IEEEauthorrefmark{2}
Quantum Labs, Cisco Systems, 3232 Nebraska Ave, Santa Monica, California 90404, USA\\
Emails: rnejabat@cisco.com, ekaur@cisco.com
}
}

\date{\today}

\begin{document}
\maketitle

\begin{abstract}

Distributed quantum computing offers a scalable alternative to monolithic quantum processors by networking smaller quantum modules through shared entangled pairs. A central challenge in this setting is that inter-module quantum operations are typically noisier than intra-module local gates, which introduces additional noise into the system. In this work, we analyze distributed lattice surgery under heterogeneous noise conditions, focusing in particular on the merge operation as one of its fundamental subroutines. Specifically, we discuss the XX merge operation between two rotated surface-code patches hosted on two different quantum processors. We characterize logical errors in the resulting H-shaped spacetime diagram and estimate thresholds using a minimum-weight perfect matching (MWPM) decoder. We use a phenomenological noise model and derive distinct bulk and seam error rates to approximate a circuit-level noise model that includes contributions from local CNOT gates, noisy entangled pairs, idle errors, and readout errors. Our results provide practical insights into selecting the optimal surface-code distance, establishing target local-gate fidelities, and determining the tolerable entangled-pair fidelity required for logical operations in a distributed architecture.

\end{abstract}



\section{Introduction}

Quantum computing is moving beyond the \textit{Noisy Intermediate-Scale Quantum (NISQ) era}, 
in which noisy qubits, rapid error accumulation, and devices containing about 50 to 1,000 qubits 
limit circuit depth and computational utility~\cite{Preskill2018,Brandhofer2021}. 
As hardware improves, larger quantum processors are beginning to support computations that go beyond 
small-scale demonstrations, a transition often discussed in terms of \textit{Quantum Utility} 
and \textit{Intermediate-Scale Quantum (ISQ)} systems~\cite{FromNISQtoISQs,pascuzzi2024quantumcentricsupercomputingphysicsresearch}. 
At the same time, scaling these computations to useful applications requires greater error protection, 
making error-corrected operation an important step toward 
\textit{Fault-Tolerant Application-Scale Quantum (FASQ)} computing~\cite{Preskill2025,eisert2025mindgapsfraughtroad}.

\vspace{2pt}


\vspace{2pt}
Despite significant progress, practical large-scale quantum computation remains limited by hardware challenges such as physical noise, limited qubit counts, crosstalk, and constrained qubit connectivity~\cite{FellousAsiani2021,Chen2023}. Addressing these limitations requires architectures that can support fault-tolerant operations, together with \textit{quantum error correction} to protect information from errors. A growing consensus is that scalable quantum computing will likely require architectural approaches beyond the monolithic devices. \textit{Distributed quantum computing (DQC)} offers one such promising alternative, in which smaller quantum processors are interconnected to form a larger computational system~\cite{Cuomo2024,Chandra2024,Kaur2025,Main2025}. In this setting, shared entangled pairs serve as a key resource for mediating nonlocal operations across module boundaries~\cite{Monroe2014,Nickerson2013,Ramette2024,Chandra2026,Jacinto2026}.

\vspace{2pt}
While shared entangled pairs enable connectivity between distant modules, they also introduce additional noise into the system~\cite{Ramette2024}. Consequently, \textit{fault-tolerant computation} across module boundaries requires an error-correcting code that can maintain a high threshold under this additional noise while remaining compatible with near-term hardware. This makes the planar surface code well suited to distributed settings, as it combines a relatively high threshold, \(\sim 1\%\) under a circuit-level noise model, with geometrically local stabilizer measurements that map naturally onto two-dimensional nearest-neighbor qubit layouts~\cite{Fowler2012}. Furthermore, its ability to tolerate additional boundary noise without substantial threshold degradation makes it a strong candidate for \textit{fault-tolerant distributed quantum computation (FT-DQC)}~\cite{Ramette2024,Larasati2025,Shalby2025}.

Although prior work on surface codes has focused primarily on \textit{memory experiments}, the emphasis has gradually shifted toward fault-tolerant computation using surface codes and other quantum error-correcting codes~\cite{Terhal2015,Katabarwa2024,Chandra_Color_Code}. This broader effort includes the development and analysis of logical gate implementations, as well as compilation strategies such as \textit{Clifford+$T$ compilation}~\cite{Litinski2019,Molavi2025}. Many architectural and compilation studies leverage the diverse toolkit available for surface-code logical operations, including transversal gates~\cite{m7tq-9v3g} and lattice surgery~\cite{Horsman2012,Vuillot2019}.

While \textit{transversal logical gates} are well suited to architectures with flexible or long-range connectivity, such as trapped-ion and neutral-atom platforms~\cite{SerraPeralta2026}, they are less compatible with superconducting-qubit architectures, where nearest-neighbor constraints make qubit-wise interactions across encoded blocks difficult. Moreover, the \textit{Eastin--Knill theorem} rules out a universal set of transversal gates for any quantum error-correcting code~\cite{Eastin2009}. These limitations make \textit{lattice surgery} an important tool for implementing logical operations on connectivity-constrained hardware. Using joint stabilizer measurements between neighboring code patches, lattice surgery enables key fault-tolerant operations, including logical CNOT gates and magic-state injection~\cite{Litinski2019,Erhard2021}.

\vspace{2pt}



In distributed architectures, lattice surgery can be adapted by replacing local joint measurements between the boundaries of surface-code patches with entanglement-mediated operations. These implementations are referred to as \textit{remote lattice surgery}~\cite{Lee2022} or \textit{distributed lattice surgery}~\cite{https://doi.org/10.48550/arxiv.2312.01246}. However, this introduces challenges absent in monolithic implementations, including the need for high-fidelity entangled (Bell) pairs, as well as additional timing and scheduling constraints. Building on this approach, recent studies have explored different challenges associated with quantum-circuit compilation in distributed architectures~\cite{Lee2022,https://doi.org/10.48550/arxiv.2312.01246,keskin2025latticesurgeryawareresource,https://doi.org/10.48550/arxiv.2603.06513}.
\vspace{2pt}

Beyond compilation, recent works have examined distributed lattice surgery from complementary operational and architectural perspectives. Ref.~\cite{sunami2025entanglementboostinglowvolumelogical} introduced the \textit{link-limited volume (LLV)} metric to quantify the operational cost of preparing high-fidelity logical Bell pairs, accounting for both Bell-pair consumption and local operations. This metric was applied to distributed lattice surgery under the assumption of noiseless \((\overline{X}\overline{X})\) and \((\overline{Z}\overline{Z})\) measurements, which correspond to merge operations. Ref.~\cite{Mrton2025} analyzed a lattice-surgery-based teleportation protocol using noisy entangled links. For monolithic architectures, Ref.~\cite{Domokos2024} characterized logical CNOT errors using a \textit{phenomenological noise model}, leaving circuit-level noise analysis for future work.

\vspace{2pt}




Motivated by these recent works and aiming to bridge the gap between \textit{monolithic and distributed architectures} for logical CNOT operations under circuit-level noise, we focus on characterizing logical errors in the \textit{merge operation}. We describe an effective phenomenological noise model informed by circuit-level considerations. This model captures the dominant error sources in a distributed setting, including Bell-pair infidelity and other local noise sources. By varying these parameters, we analyze the impact of Bell-pair noise on logical error rates and the thresholds of the merge operation.

\vspace{2pt}
The main contributions of this work are as follows:

\begin{itemize}[leftmargin=1.2em]
 \item We consider a phenomenological noise model for two rotated surface codes hosted on distinct quantum processors connected by Bell pairs. Based on this setup, we derive effective error probabilities for bulk and seam qubits by incorporating noise from the imperfect Bell pairs, CNOT gates, readout, and idling.

    \vspace{4pt}

    \item We characterize logical \(Z\) errors in the \((\overline{X}\overline{X})\) merge operation between two distance-\(d\) rotated surface codes using an \textit{H-shaped} spacetime diagram. Applying minimum-weight perfect matching (MWPM) decoder, we compute logical error rates as they vary with the physical error rate and estimate the merge operation threshold.

    \vspace{4pt}

    \item We quantify the effect of Bell-pair noise by introducing a scaling factor \(k \in \{1,3,5,7,9,11\}\) relative to other noise sources described by a uniform physical error rate. By systematically increasing this parameter, we analyze its impact on both the logical error rate and the threshold.
\end{itemize}








This paper is organized as follows: In Sec.~\ref{BG}, we review the relevant background theory. In Sec.~\ref{Distributed_Setup}, we describe the merge operation in a distributed setup comprising two quantum processors and discuss the associated noise model. In Sec.~\ref{results}, we present our results on the distributed merge operation, focusing on how logical error rates and thresholds vary under elevated entanglement noise. Finally, in Sec.~\ref{conclusion}, we summarize our findings and outline directions for future work.

\section{Background}\label{BG}

In this section, we briefly review the background theory underlying this work, with a focus on the rotated surface code, logical errors, and lattice surgery.

\subsection{Rotated Surface Code}

\begin{figure}[h!]
    \centering
    \includegraphics[width=0.85\columnwidth]{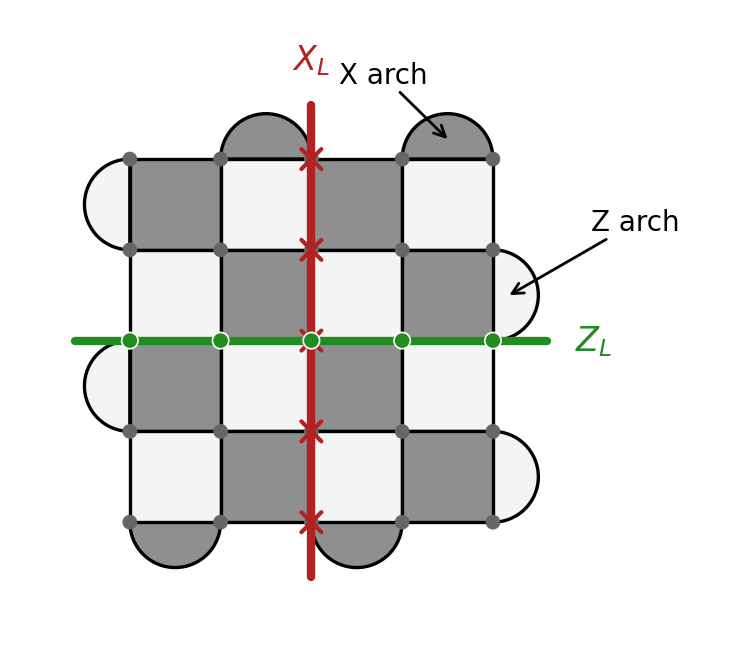}
    \caption{Rotated Surface Code for $d=5$. The gray ($X$-type) and white ($Z$-) squares represent the plaquettes (stabilizer generators) of the code. Data qubits are located at the vertices (grey circles). The vertical line indicates the logical X operator ($X_L$), and the horizontal line indicates the logical Z operator ($Z_L$). The X and Z arches, indicated by arrows, represent the stabilizers at the boundary. We use red for $X_L$ and Green $Z_L$ to represent the logical operators.}
    \label{fig:surface_code_patch}
\end{figure}

The rotated surface code is a planar CSS stabilizer code with open boundaries~\cite{Tomita2014} (see Fig.~\ref{fig:surface_code_patch}). A distance-\(d\) rotated patch encodes one logical qubit (\(k=1\)) using \(n=d^2\) data qubits,
\begin{equation}
[[n,k,d]] = [[d^2,1,d]].
\end{equation}

The stabilizer group \(\mathcal{S}\) is generated by local \(X\)-type and \(Z\)-type plaquette stabilizers,
\begin{equation}
\begin{aligned}
\mathcal{S} &= \langle A_f, B_g : f\in F_X,\; g\in F_Z \rangle,\\
A_f &= \prod_{i\in V(f)} X_i,\qquad
B_g = \prod_{i\in V(g)} Z_i.
\end{aligned}
\end{equation}
Here, \(A_f\) and \(B_g\) represent the \(X\)-type and \(Z\)-type plaquette stabilizers associated with the sets of plaquettes \(F_X\) and \(F_Z\), respectively, and \(V(\cdot)\) denotes the data qubits on a given plaquette. Bulk stabilizers have weight four, while boundary stabilizers have weight two. The codespace satisfies
\begin{equation}
A_f \ket{\psi_L}=\ket{\psi_L}, \qquad B_g \ket{\psi_L}=\ket{\psi_L},
\end{equation}
for all \(f\in F_X\) and \(g\in F_Z\), where \(\ket{\psi_L}\) represents a logical state in the codespace.


\subsection{Minimum-Weight Perfect Matching Decoding}


For the rotated surface code, decoding is commonly performed using minimum weight perfect matching (MWPM)~\cite{Dennis2002,Higgott2022}. Since the code is CSS, \(X\) and \(Z\) errors are decoded independently: \(Z\) errors are detected by \(X\)-type stabilizers, while \(X\) errors are detected by \(Z\)-type stabilizers. The patch has two pairs of opposite boundaries, conventionally labeled as \(X\)-type and \(Z\)-type boundaries, and logical operators correspond to nontrivial Pauli strings connecting the  boundaries of same type. 
\vspace{4pt}

We note that in the surface-code literature, the conventions used to label \(X\)- and \(Z\)-type edges (or boundaries) are not uniform across references. In some works, an edge is labeled by the type of stabilizer associated with it, whereas in others it is labeled by the logical operator supported on or measured along it. In this work, we follow the convention of Ref.~\cite{kottmann2025latticesurgery} for describing edge or boundary. Under this convention, edges are named according to the logical operators they support, rather than the type of arches used to depict them (see Fig.~\ref{fig:surface_code_patch}). Accordingly, an edge drawn with \(X\) arches is called a \(Z\) edge because measuring along it realizes the logical \(\overline{Z}\) operator, whereas an edge drawn with \(Z\) arches is called an \(X\) edge (See Fig.~\ref{fig:surface_code_patch}). 


To account for errors on both data qubits and syndrome measurements, stabilizers are measured repeatedly and MWPM decodes changes in the syndrome between consecutive rounds. If \(s_f(t)\in\{0,1\}\) is the outcome of stabilizer \(f\) in round \(t\), with \(0\) and \(1\) corresponding to eigenvalues \(+1\) and \(-1\), respectively, then a detection event is defined as~\cite{Fowler2012proof},
\begin{equation}
\delta_f(t)=s_f(t)\oplus s_f(t-1).
\end{equation}
Thus, \(\delta_f(t)=1\) marks a change in the measured syndrome.

The detection events form the vertices of the decoding graph, with boundary vertices included to represent error chains ending at code boundaries. Edges represent error mechanisms or effective error chains that can produce the corresponding detection event pattern, and are assigned the standard log-likelihood weight,
\begin{equation}
w_e=\log\!\left(\frac{1-p_e}{p_e}\right),
\end{equation}
where \(p_e\) is the associated error probability. MWPM finds a minimum weight matching of the observed detection events and infers a recovery chain \(R\). If \(E\) is the actual error chain, decoding succeeds when \(ER\) is homologically trivial and fails when \(ER\) is equivalent to a nontrivial logical operator~\cite{Higgott2022}.

\subsection{Nonlocal CNOT Gate}

\begin{figure}[h!]
    \centering
    \includegraphics[width=0.48\textwidth]{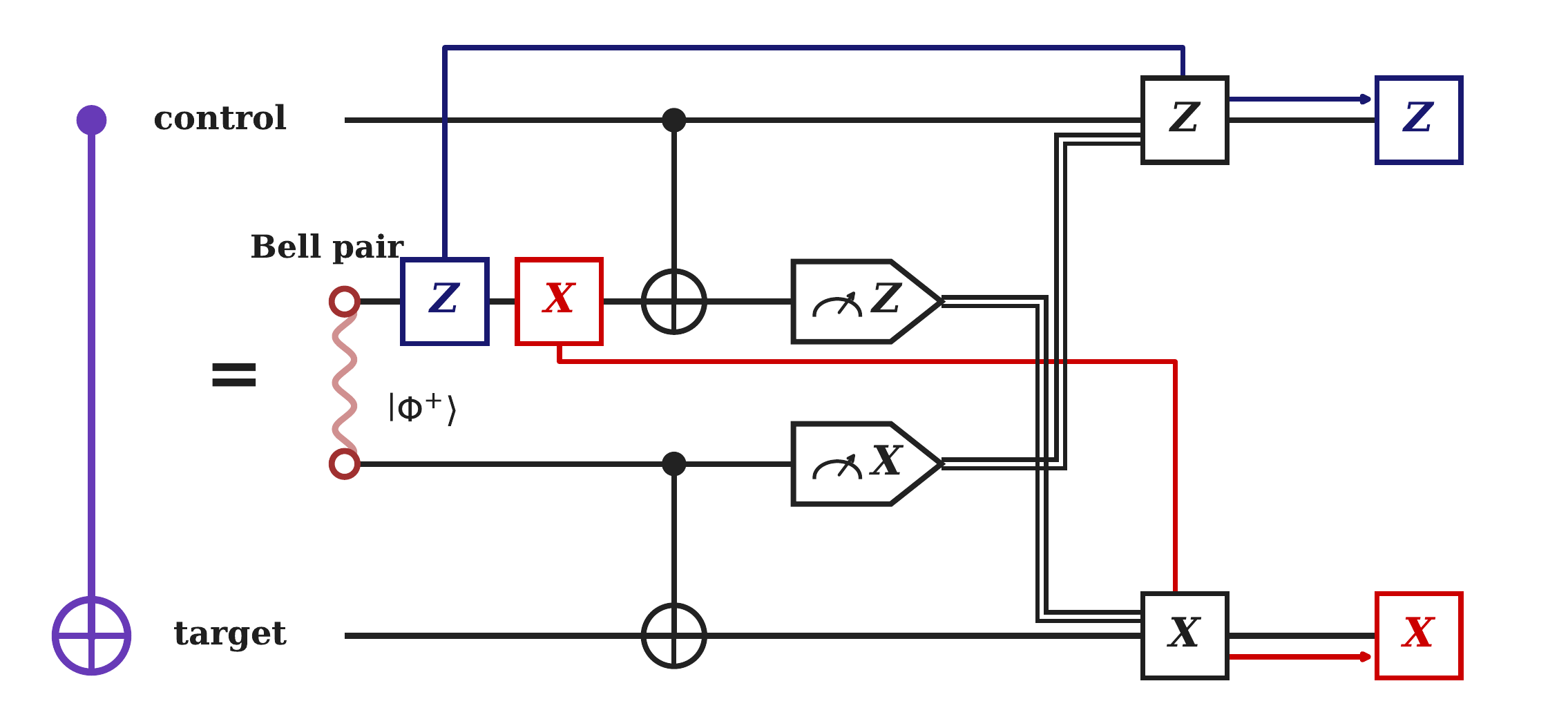}
    \caption{
        Entanglement-assisted  nonlocal CNOT gate. A shared Bell pair is used to mediate the interaction between spatially separated control and target qubits, followed by local CNOTs, measurements, and classical feedforward for Pauli corrections.
    }
    \label{fig:nonlocal_cnot_decomposition}
\end{figure}

We can use an entangled pair as the resource for implementing a nonlocal CNOT between qubits in two different QPUs. This operation can be performed using local CNOTs, measurements of the Bell-pair qubits, and classical feedforward for Pauli corrections (See Fig.~\ref{fig:nonlocal_cnot_decomposition}). However, this introduces additional noise into the system, which we consider in our noise model. In particular, phase-flip errors propagate to the control qubit, while bit-flip errors propagate to the target qubit~\cite{Ramette2024}. Thus, the fidelity of the entangled pair plays a key role in determining the performance of logical operations.

\subsection{Lattice Surgery}

Lattice surgery is a fault-tolerant method for implementing logical operations between neighboring surface-code patches using joint stabilizer measurements~\cite{Horsman2012}. Instead of applying transversal gates between corresponding physical qubits, we temporarily modify the stabilizers along a shared boundary to measure a joint logical parity. If two patches are joined along compatible boundaries, the interface stabilizers \(\{S_f^{\mathrm{int}}\}\) are chosen such that,
\begin{equation}
\prod_{f\in \mathcal{I}} S_f^{\mathrm{int}} = M,
\qquad
M \in \{\overline{Z}_1\overline{Z}_2,\; \overline{X}_1\overline{X}_2\},
\end{equation}
where \(\mathcal{I}\) denotes the set of interface plaquettes along the boundary and the subscripts \(1\) and \(2\) refer to the first and second surface-code patches, respectively. Lattice surgery consists of two elementary operations: a merge, which joins two patches by activating interface stabilizers to measure a joint logical operator, and a split, which separates them by removing those checks and restoring independent logical patches.

\vspace{2pt}
\emph{Merge.}
In a merge operation, two adjacent patches are combined into a single, temporarily larger code block. The boundary checks along the shared interface are replaced by new interface stabilizers, and syndrome extraction is repeated for \(d\) rounds to preserve fault tolerance. After decoding, the measurement yields an outcome \(m\in\{\pm1\}\), corresponding to the projection,
\begin{equation}
\Pi_m = \frac{1}{2}(I + mM).
\end{equation}
Thus, the merge measures the logical parity \(M\) while temporarily reducing the number of encoded degrees of freedom by one.

\vspace{4pt}
\emph{Split.}
A split operation divides a single extended patch into two independent patches by terminating the joint stabilizer measurements along a chosen interface and subsequently treating the two halves as distinct logical entities. This process maps the logical operators of the extended patch onto the two resulting patches. To ensure fault tolerance, both patches undergo \(d\) rounds of independent syndrome extraction. This temporal overhead provides the necessary syndrome history to distinguish physical data qubit errors from measurement errors.



\begin{figure}[htbp]
    \centering
    \includegraphics[width=0.9\linewidth]{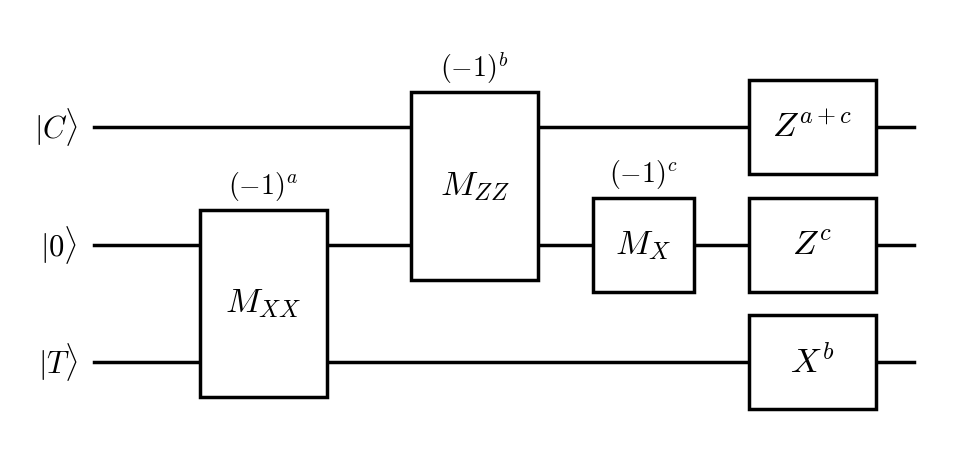}
    \caption{Lattice surgery circuit for a logical CNOT gate. The control ($|C\rangle$) and target ($|T\rangle$) patches interact via an ancilla ($|0\rangle$) using sequential joint parity measurements. The final byproduct Pauli corrections are dictated by the measurement outcomes $a, b,$ and $c$.}
    \label{fig:lattice_surgery_cnot}
\end{figure}

\vspace{4pt}
\emph{Logical CNOT.}
A logical CNOT between a control patch \(C\) and a target patch \(T\) can be implemented by combining merge and split operations with an intermediate ancilla patch \(A\) prepared in the logical state \(\ket{0}_L\)~\cite{Vuillot2019}. As illustrated in Fig.~\ref{fig:lattice_surgery_cnot}, the protocol sequentially measures the joint logical parities,
\begin{equation}
M_{XX} = \overline{X}_A\overline{X}_T,
\qquad
M_{ZZ} = \overline{Z}_C\overline{Z}_A,
\end{equation}
yielding outcomes \((-1)^a\) and \((-1)^b\), respectively. This is followed by a measurement of the ancilla in the logical \(X\)-basis, yielding outcome \((-1)^c\). These three measurement outcomes determine the necessary Pauli-corrections leading to a net logical action of \(\mathrm{CNOT}_{C\to T}\).



\section{Characterizing Logical Errors in Merge Operation in a Distributed Setting}\label{Distributed_Setup}

Lattice surgery requires surface code patches to be close enough for a contiguous bridge of coupling qubits to be initialized and used in joint stabilizer measurements. This local connectivity requirement can make scaling large quantum processors challenging. Remote, or distributed lattice surgery circumvents this limitation by enabling merge operations between physically separated logical qubits encoded in surface code patches~\cite{Lee2022,https://doi.org/10.48550/arxiv.2603.06513}. The protocol consumes shared Bell pairs between remote quantum processors to enable the bridge stabilizers. Boundary qubits from each patch interact locally with their respective halves of the shared Bell pairs, and the joint stabilizer parity is inferred through classical communication.

\vspace{2pt}

While decoupling connectivity from the physical hardware layout offers substantial architectural flexibility, it changes the error propagation profile of the logical operation. In local lattice surgery, fault-tolerance is achieved by repeating stabilizer measurements over multiple rounds, where the dominant noise sources are local two-qubit gate infidelities and measurement errors. Distributed lattice surgery introduces additional noise from non-local gate operations. The generation, distribution, and consumption of shared entanglement introduce additional noise sources, including Bell pair infidelities, local gate operations and measurements, and memory decoherence. Consequently, standard assumptions of uniform, independent depolarizing noise across the patch are insufficient to accurately capture the noise in distributed lattice surgery.

\vspace{2pt}
Previous studies have evaluated the logical error rates and fault-tolerance thresholds of different operations involving lattice surgery; however, these investigations have predominantly relied on standard phenomenological models for monolithic architectures~\cite{Domokos2024} or have assumed the merge process to be noiseless~\cite{sunami2025entanglementboostinglowvolumelogical}. To accurately assess the feasibility of distributed lattice surgery, it is necessary to employ a noise model that captures the additional noise introduced by  nonlocal gates. To address this gap, we consider a noise model that explicitly incorporates Bell pair noise and other noise sources in nonlocal gate operations. By simulating the distributed merge operation under these noise conditions, we evaluate the resulting logical error rates and thresholds, thereby providing a more realistic assessment of merge operation in a distributed setting.


\subsection{The Merge Operation}
\begin{figure}[h!]
    \centering
    \includegraphics[width=0.45\textwidth]{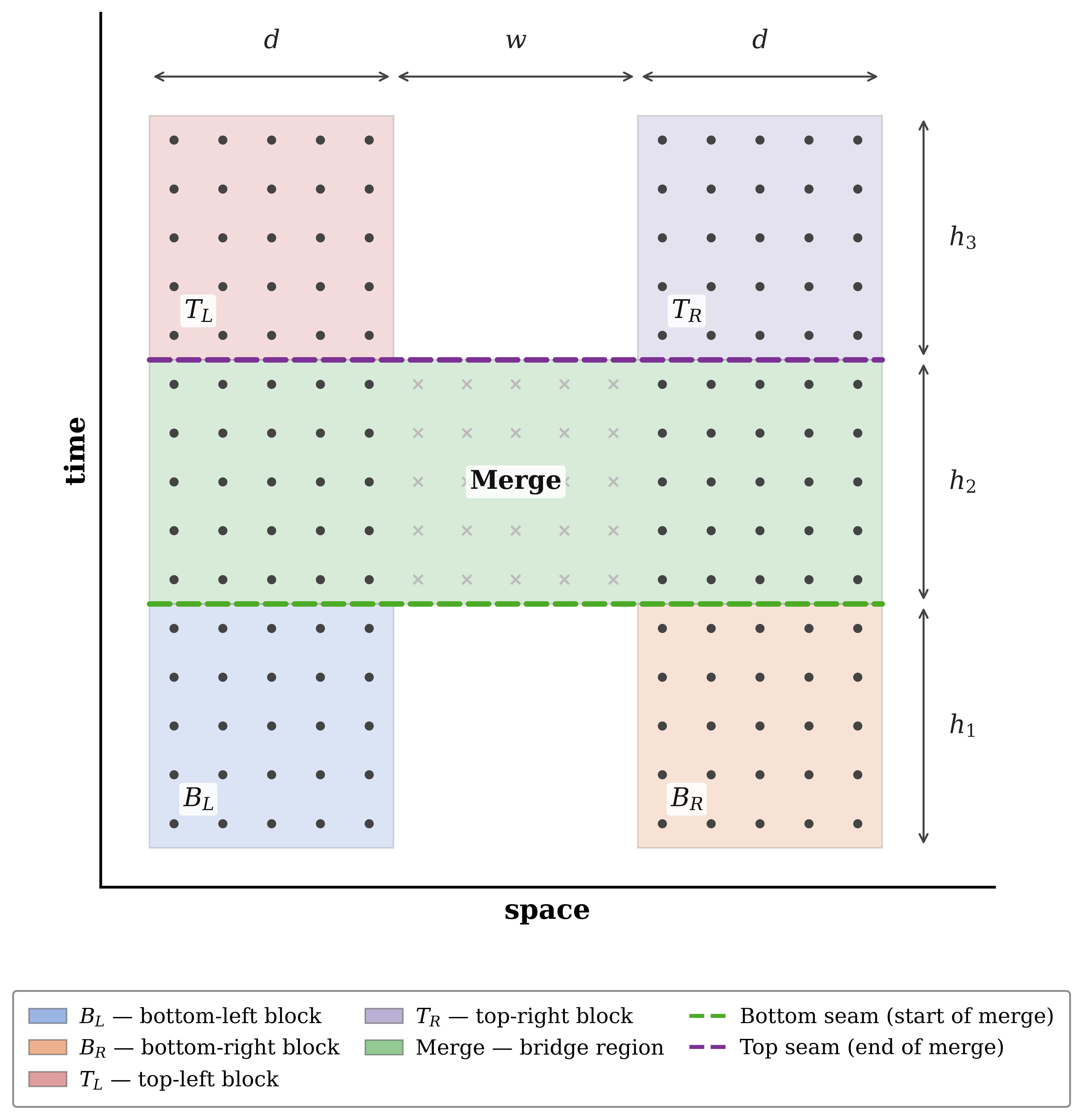}
    \caption{%
    Spacetime diagram of the \textit{H-shaped} geometry for the merge operation in lattice surgery. The horizontal and vertical axes represent the qubits and syndrome extraction rounds, respectively.
    }
    \label{fig:H-geometry}
\end{figure}


The spacetime representation provides a useful way to visualize the merge operation. As illustrated in Fig.~\ref{fig:H-geometry}, merge operation traces out an \textit{H-shaped} geometry in spacetime diagram. The figure shows a two dimensional projection of the three dimensional spacetime diagram (See Fig.~\ref{fig:3d_geometry}). The horizontal axis represents the spatial layout of the physical qubits, while the vertical axis denotes time, measured in rounds of syndrome extraction.

\vspace{2pt}


The spacetime diagram is divided into five regions. Before the merge begins, syndrome extraction is performed separately on the two surface code patches for \(h_1\) rounds, forming the bottom-left block \(B_L\) and the bottom-right block \(B_R\). The coupling qubits between the two patches, marked by \(\times\), are inactive during this initial stage.

\vspace{2pt}

At time \(t=h_1\), the merge is initiated along the \emph{bottom seam}, shown by the green dashed line. During the following \(h_2\) syndrome extraction rounds, the coupling qubits are activated and used to measure the joint interface stabilizers. This joins the two patches into a single enlarged surface code patch. The resulting bridge region extends across both patches and the gap between them.

\vspace{2pt}

Finally, at time \(t=h_1+h_2\), the split operation occurs along the \emph{top seam}, shown by the purple dashed line. The interface stabilizers are deactivated, and syndrome extraction is performed on the two independent patches for \(h_3\) post merge rounds, forming the top-left block \(T_L\) and the top-right block \(T_R\).
\vspace{2pt}



\begin{figure}[h!]
    \centering
    \includegraphics[width=0.5\textwidth]{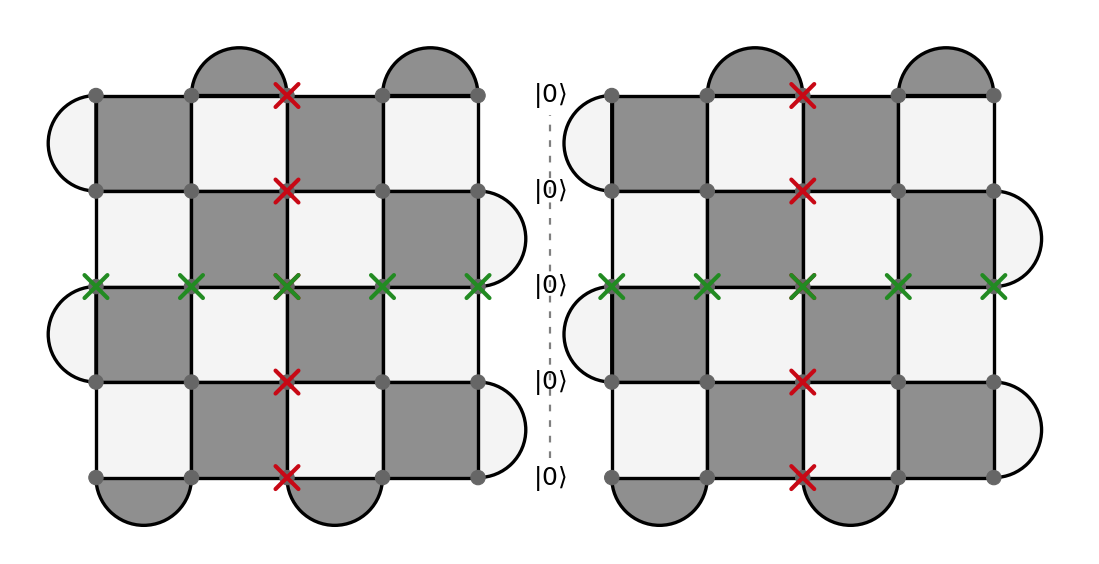}
    \caption{Ancilla mediated merge operation between two \(d=5\) rotated surface code patches. The central column of qubits, initialized in \(\ket{0}\), forms the bridge region used to couple the two patches during the merge. The dashed line marks the interface where joint stabilizers are measured, enabling a logical parity measurement between the two encoded qubits.}
    \label{fig:merge_operation}
\end{figure}

\vspace{2pt}
The \(\overline{X}\overline{X}\) measurement is implemented through three main steps (See Figs. \ref{fig:H-geometry} and~\ref{fig:merge_operation}):
\begin{itemize}
    \item The coupling data qubits in the bridge region are first prepared in the \(\ket{0}\) state.
    \vspace{2pt}
    
    \item The two separate patches are then joined into an extended rectangular patch by measuring the stabilizers of the combined surface code.
    \vspace{2pt}
    
    \item The split is performed by measuring the coupling qubits in the \(Z\) basis.
\end{itemize}

\vspace{2pt}



In an ideal noiseless setting, the logical \(\overline{X}\overline{X}\) measurement outcome is obtained from the product of the newly introduced intermediate \(X\)-stabilizer measurement outcomes. In the presence of noise, however, a single round of intermediate stabilizer measurements is not sufficient, since measurement errors can corrupt the inferred \(\overline{X}\overline{X}\) outcome. To maintain fault tolerance, the merge region must be measured for at least,
\begin{equation}
    h_2 = d
\end{equation}
rounds between the merge and split operations, as shown in Fig.~\ref{fig:H-geometry}.

The merge process possesses a distinctive spacetime in three-dimensions (See Fig.~\ref{fig:3d_geometry}). The spacetime diagram of the \(\overline{X}\overline{X}\) measurement exhibits perfect spacelike boundaries at both the top and bottom, with timelike boundaries positioned in the middle. Geometrically, this structure comprises four disconnected \(X\)-edges and two disconnected \(Z\)-edges~\cite{Domokos2024}.



\vspace{2pt}

To correct errors in this operation, syndrome history is collected from the entire \textit{H-shaped} spacetime region, spanning the pre-merge phase, the bridge region, and the post-merge phase. This complete syndrome history is then passed to a decoder, such as the MWPM decoder used in this work, which processes the full three-dimensional syndrome history.



\begin{figure*}[h!]  
    \centering
    \includegraphics[width=0.8\textwidth]{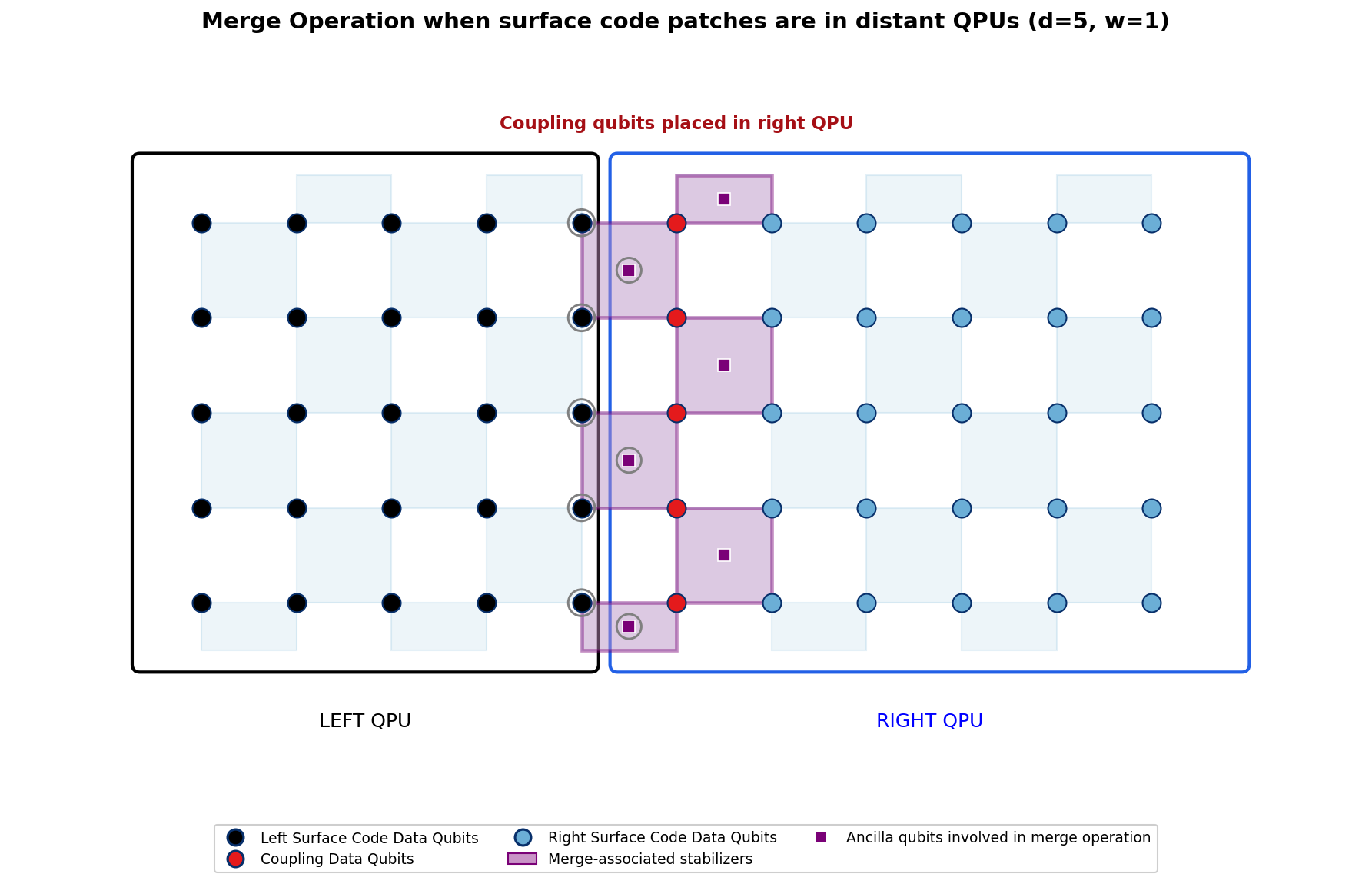}
    \caption{Visualization of the bridge step joining two rotated surface-code patches during an \(\overline{X}\overline{X}\) merge operation across different QPUs. To connect the patches, new merge-associated stabilizers (purple) and coupling data qubits (red) are introduced along the shared boundary. In our specific setup, the coupling qubits are hosted within the right QPU. Consequently, boundary checks and data interactions across the QPUs are nonlocal, requiring entangled pairs to implement the necessary nonlocal gates.}
    \label{fig:merge_operation_QPU}
\end{figure*}

\vspace{0.5cm}

\subsection{Noise Model for Distributed Merge Operation}


We consider a distributed setting in which each rotated surface code patch is hosted on a separate quantum processing unit (QPU). The two QPUs are spatially separated, with coupling qubits placed in the right QPU, as shown in Fig.~\ref{fig:merge_operation_QPU}. Consequently, a merge operation across the interface cannot be implemented using only local gates. Instead, stabilizer measurements that span the two QPUs are mediated by shared Bell pairs to perform nonlocal gate operations.

\vspace{4pt}

This leads to a spatially inhomogeneous noise model, since qubits participating in cross-QPU syndrome extraction are subject to errors associated with nonlocal operations, which differ from the errors affecting qubits involved only in intra-QPU operations. As illustrated in Fig.~\ref{fig:merge_operation_QPU}, the circled qubits, namely the data qubits of the left surface code patch and the ancilla qubits associated with the merge, reside on different QPUs. Accordingly, we distinguish four qubit classes: bulk data qubits, bulk ancilla, or syndrome, qubits, seam data qubits, and seam ancilla, or syndrome, qubits that mediate the logical coupling across the interface.

\vspace{4pt}



We map the underlying noise arising from various physical sources onto effective phenomenological error rates for each region. A phenomenological noise model consists of two independent error mechanisms. Each data qubit suffers a Pauli error with probability $p$ before each round of syndrome extraction, called as the data error rate.  Each syndrome measurement is flipped independently with probability $q$, called as the measurement (syndrome) error rate. 

\vspace{2pt}

Let $p_{\text{bulk}}$ and $q_{\text{bulk}}$ denote the effective data qubit error rate and measurement qubit error rate, respectively, for the qubits located within the QPU (both left and right). We define $p_{\text{seam}}$ and $q_{\text{seam}}$ as the effective data and measurement error rates at the seam (the QPU boundary), which have additional noise due to their involvement in non-local operations. We derive the analytical expressions for these four effective noise parameters arising from different error sources such as local gate operations, noisy Bell pairs, idle errors and measurement errors (See Appendix~\ref{Noisee Model}). We can describe this noise model with the following error rate for phase-flip (Z) noise,


\begin{subequations}\label{eq:x_error_rates}
\begin{align}
p_{\mathrm{bulk}} &=
2\varepsilon_{\mathrm{cx}} + \varepsilon_{\mathrm{idle}}, \\
q_{\mathrm{bulk}} &=
2\varepsilon_{\mathrm{cx}} + \varepsilon_{\mathrm{m}} + \varepsilon_{\mathrm{idle}}, \\
p_{\mathrm{seam}} &=
0.5\,\varepsilon_{\mathrm{B}} + 2.5\,\varepsilon_{\mathrm{cx}} + \varepsilon_{\mathrm{m}} + \varepsilon_{\mathrm{idle}}, \\
q_{\mathrm{seam}}&=
\varepsilon_{\mathrm{B}}
    + 3\varepsilon_{\mathrm{cx}}
    + 3\varepsilon_{\mathrm{m}}
    + \varepsilon_{\mathrm{idle}}.
\end{align}
\end{subequations}

where $\varepsilon_{\mathrm{cx}}$ denotes the local two-qubit gate error, $\varepsilon_{\mathrm{B}}$ represents the error due to noisy Bell pairs, $\varepsilon_{\mathrm{m}}$ is the measurement (readout) error, and $\varepsilon_{\mathrm{idle}}$ corresponds to the error due to idle noise.


We consider a uniform noise scenario with elevated Bell pair noise,
\begin{equation}
\varepsilon_{\mathrm{cx}} = \varepsilon_{m} = \varepsilon_{\mathrm{idle}} = \varepsilon,
\qquad
\varepsilon_{B} = k\varepsilon.
\end{equation}

Specifically, we take the two-qubit gate, measurement, and idle error rates to be equal to a common physical error rate $(\varepsilon)$, while the error associated with Bell pairs is assumed to be larger by a factor of $k\geq1$. Under these assumptions, the effective error probabilities can be expressed in terms of the single physical error rate $\varepsilon$ and the Bell-pair noise scaling factor $(k)$. This parametrization makes it convenient to study how additional Bell-pair noise affects the logical error rate and the threshold of the logical \textit{merge operation}.



\vspace{2pt}
Consequently, the effective error rates for the different qubits simplify to,

\begin{align}
    p_{\mathrm{bulk}} &= 3\varepsilon, \\
    q_{\mathrm{bulk}} &= 4\varepsilon, \\
    p_{\mathrm{seam}} &= \frac{\varepsilon}{2}(k+9), \\
    q_{\mathrm{seam}} &= \varepsilon(k+7).
\end{align}

\subsection{Decoding errors in the spacetime diagram of the
\texorpdfstring{\(XX\)}{XX} merge operation}


To decode the logical errors in \(\overline{X}\overline{X}\) merge operation, we construct a 3D spacetime syndrome graph from repeated \(X\)-stabilizer measurements. Data-qubit errors are assigned to spacelike edges, while measurement errors between consecutive rounds are assigned to timelike edges. Since \(X\)-type checks detect \(Z\)-type errors, we use these checks to decode \(Z\)-error strings. We apply minimum-weight perfect matching (MWPM) to determine the most probable correction string, with spacelike and timelike edges weighted according to the phenomenological error probabilities derived from our noise model.

\vspace{2pt}

\begin{figure}[htbp]
    \centering
    \includegraphics[width=0.35\textwidth]{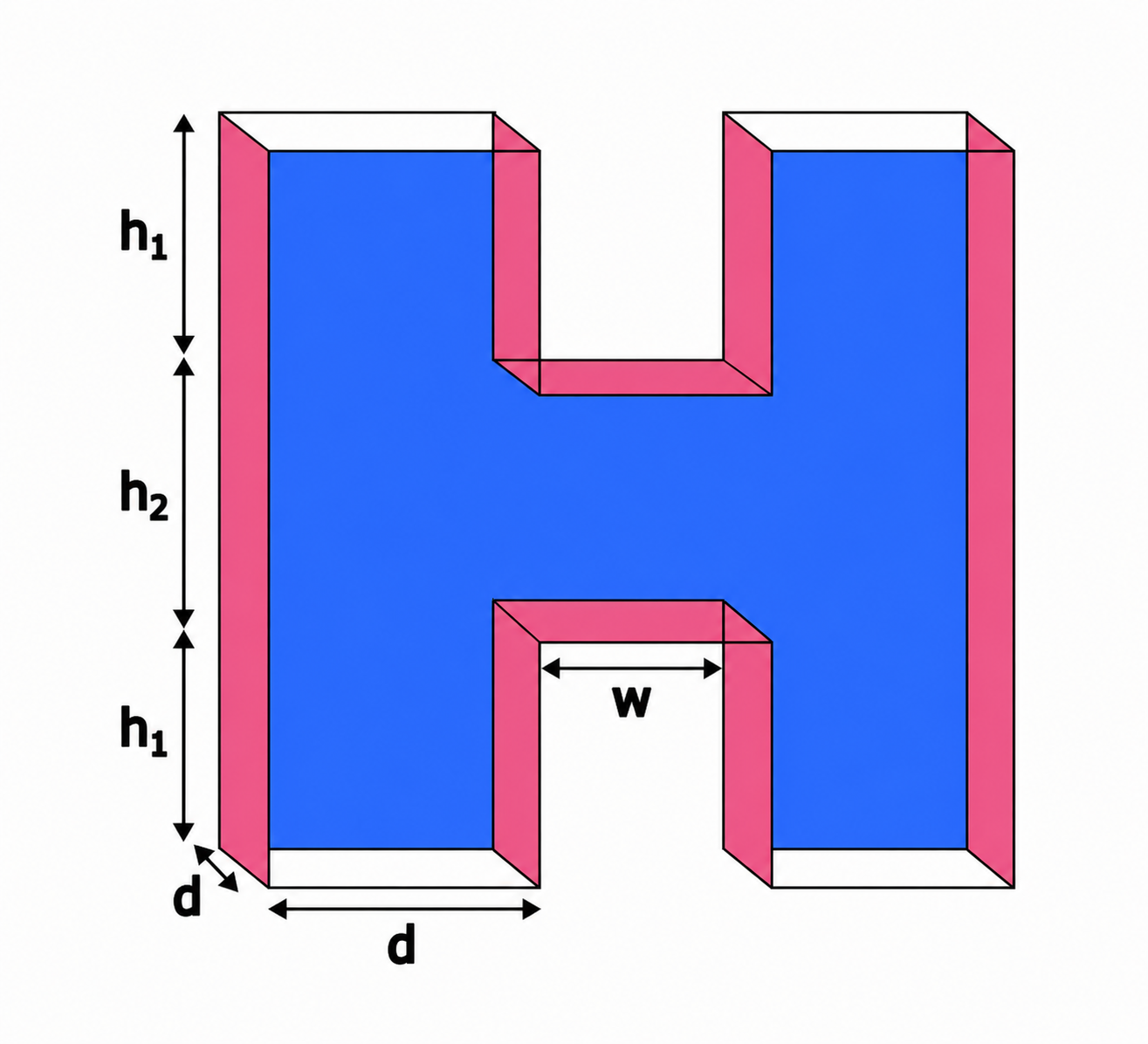}
    \caption{ The \(XX\) merge operation contains four disconnected \(X\)-type boundaries and two
\(Z\)-type boundaries. In the figure, the red regions indicate the \(X\)-type
boundaries, while the blue regions indicate the visible \(Z\)-type boundary. The
second \(Z\)-type boundary is hidden by the perspective of the layout. The
geometry is specified by the surface-code distance \(d\) and the bridge width
\(w\) connecting the two surface-code patches. The parameter \(h_1\) denotes the
number of measurement rounds before and after the merge, while \(h_2\) denotes
the number of measurement rounds during the interval in which the two patches are
merged. This schematic follows the layout of the spacetime diagram in Ref.~\cite{Domokos2024}.}
    \label{fig:3d_geometry}
\end{figure}

Using the MWPM decoder, we compute the residual error string by adding the predicted correction to the actual error modulo two. To determine whether a logical error has occurred, we evaluate the parity of this residual string along the four disconnected $X$-type boundaries of our H-shaped spacetime volume: the left, right, top-middle, and bottom-middle boundaries (See Fig.~\ref{fig:3d_geometry}). We classify a decoding cycle as successful if and only if the parity across all four boundaries evaluates to zero. If we detect a non-zero parity on any of these boundaries, it indicates that the residual string is homologically non-trivial, and we record the run as a logical failure. This is based on the topological principle that logical errors manifest as syndrome-free strings connecting disconnected boundaries of the same type~\cite{Domokos2024}.

\section{Results and Discussion}\label{results}

To evaluate the performance of the distributed $\overline{XX}$ merge operation, we characterize the logical error rate of $Z$ errors, and the fault-tolerance threshold, $\varepsilon_c$, as a function of the Bell-pair noise scaling factor, $k \in \{1, 3, 5, 7, 9, 11\}$. These results were obtained using 200,000 Monte Carlo shots per data point, varying the code distance across $d \in \{5, 7, 9, 11, 13\}$ with $w = 1$, and $h_{1}=h_{2} = d$.


\begin{figure*}[h!]

  \centering
  \includegraphics[width=0.95\textwidth]{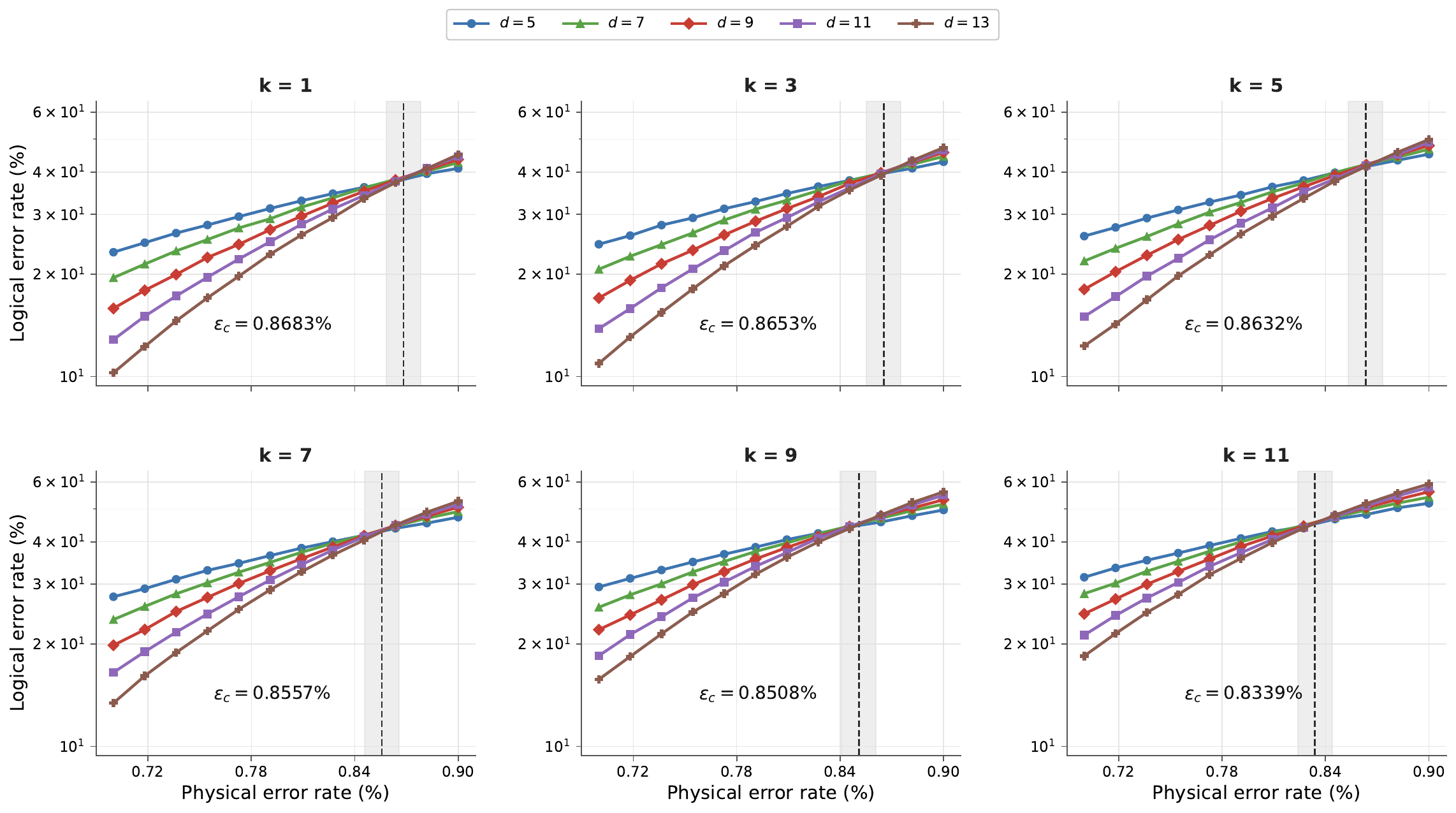}
  \caption{Logical error rate versus base physical error rate $\varepsilon$ for distances
$d=5,7,9,11,13$ using six values of the Bell-pair noise scale
$\varepsilon_B = k\,\varepsilon$. Colors and markers distinguish the code
distances, as indicated in the legend. Each panel corresponds to a fixed value
of $k$. The dashed vertical line marks the estimated threshold
$\varepsilon_c$, and the shaded band shows the fixed interval
$[\varepsilon_c-0.010\%,\,\varepsilon_c+0.010\%]$ around the threshold.}
  \label{fig:threshold_curves}
\end{figure*}

\begin{figure}[h!]
  \centering
  \includegraphics[width=0.5\textwidth]{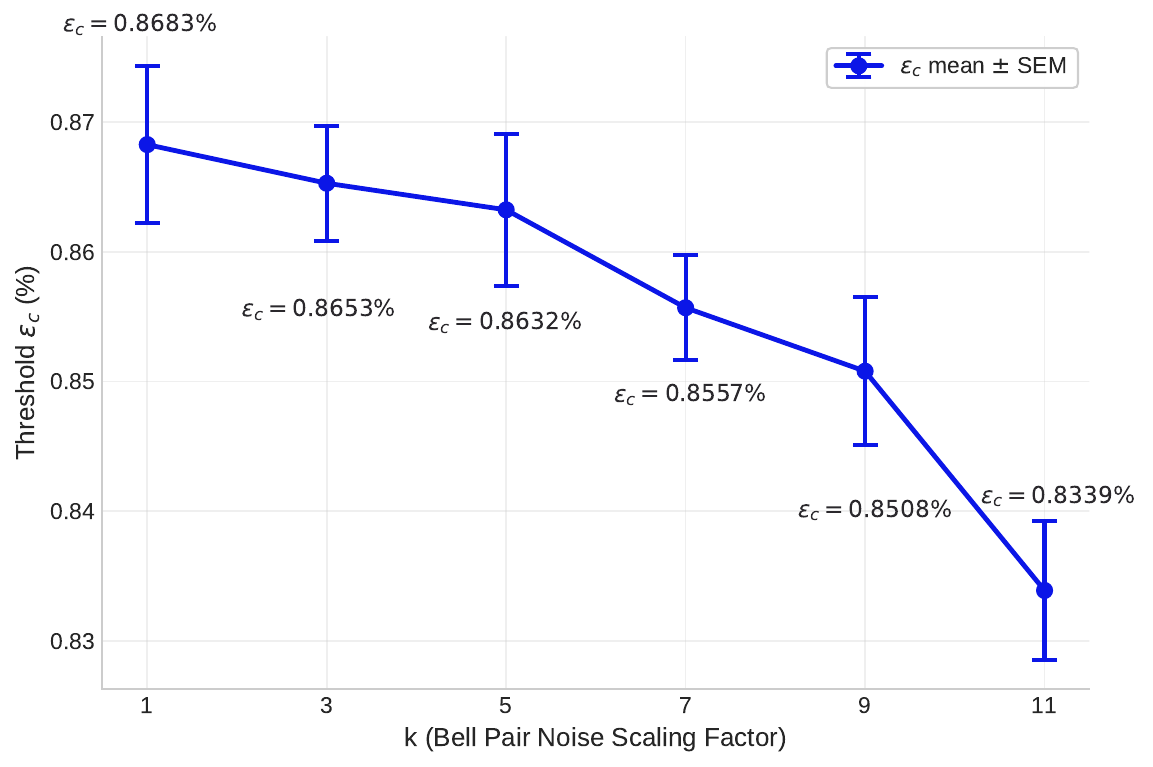}
  \caption{ Logical error threshold $\varepsilon_c$ as a function of the Bell pair noise scaling factor $k$. Each point represents the mean threshold extracted from linear sign-change interpolation across four adjacent-distance crossing pairs $(d_i, d_{i+1})$ for $d \in \{5, 7, 9, 11, 13\}$, with error bars indicating the standard error of the mean (SEM). 
    }
  \label{fig:error_plot}
\end{figure}

\vspace{2pt}

Fig.~\ref{fig:threshold_curves} illustrates the logical error rates as a function of the physical error rate, $\varepsilon$, for each value of $k$. The crossing point of the distance curves indicates the threshold, $\varepsilon_c$, marked by the vertical dashed lines with a small fixed region around the threshold. As the Bell-pair noise factor $k$ increases, these threshold markers shift slightly leftward toward lower physical error rates, indicating a reduction in the overall error tolerance due to the noisier entanglement pairs.

\vspace{2pt}

Fig.~\ref{fig:error_plot} illustrates this threshold degradation, showing a monotonic decrease from $\varepsilon_c = 0.8683\%$ at $k=1$ to $\varepsilon_c = 0.8339\%$ at $k=11$. Despite the Bell-pair error rate increasing by a factor of 11 relative to the local gate errors, the threshold undergoes an absolute reduction of approximately $0.034\%$. These results suggest that  logical \textit{merge operation} can operate with noisy entanglement links while maintaining a stable fault-tolerance threshold.

\vspace{2pt}
Our results highlight two important observations. First, the observed threshold of \(\varepsilon_c \approx 0.86\%\) at \(k=1\) lies within the range reported in existing circuit-level thresholds of the rotated surface code, where thresholds generally fall between \(0.5\%\) and \(1.0\%\), depending on the implementation and decoder used~\cite{ORourke2025,Higgott2023}. Second, even for \(k=11\), where the Bell pair noise is increased by an order of magnitude relative to \(k=1\), the threshold reduction remains modest. We explain the reasons for these observations in the following paragraphs.

\vspace{2pt}

Domokos \textit{et al.}~\cite{Domokos2024} investigated a monolithic lattice-surgery based CNOT operation under a standard phenomenological noise model, and found thresholds of \(3\%\) for independent noise and \(4.5\%\) for depolarizing noise. The authors suggested that these thresholds should apply broadly to any lattice-surgery based logical operations, including under circuit-level noise models. Their reasoning was supported by scaling arguments~\cite{Wang2003}: in the large code distance limit, \(d \to \infty\), the statistical mechanical model governing the decoding procedure is dominated by the bulk rather than boundary details. Using a noise model  to approximate a circuit-level noise model, we numerically \textit{verified} this behavior. It is important to note that, although the threshold remains close to that of standard memory experiments, the more complex spacetime structure of these logical operations introduces additional error classes, leading to lower logical error rates than those observed in standard syndrome extraction experiments.



\vspace{2pt}

We also find that increasing the Bell-pair-induced seam noise up to a scaling factor of \(k=11\) causes only a \textit{modest reduction} in the threshold of the merge operation. This behavior is consistent with prior studies on surface-code memory studies. For example, Ramette \textit{et al.}~\cite{Ramette2024} demonstrated that boundary regions can tolerate noise roughly an order of magnitude higher than the bulk. Similar scaling arguments explain our results: the bulk contains \(\mathcal{O}(d^2)\) qubits per syndrome extraction round, whereas the seam contains only \(\mathcal{O}(d)\) qubits. In the limit of large code distance (\(d\to\infty\)), the seam therefore contributes a vanishing fraction of the total spacetime error volume; thus, the asymptotic threshold is dominated by the bulk noise.

Although the spacetime geometry of the merge operation is more complex than that of a memory experiment and can support additional error string classes, our results indicate that the same scaling intuition continues to hold for the merge operation for surface code. Based on the same asymptotic argument, we expect similar boundary-noise tolerance to persist for larger logical operations built from merge and split primitives.

\vspace{2pt}

Our results demonstrate that logical operations, such as the merge operation, enable systems to accommodate noisier interconnects. This translates directly into a favorable \textit{entanglement rate-fidelity tradeoff}: entanglement distillation consumes many Bell pairs and operates probabilistically, thereby severely degrading the effective entanglement rate. By allowing lower-fidelity pairs and minimizing distillation steps, systems can achieve significantly higher generation rates. Ultimately, this flexibility eases the stringent entanglement requirements for fault-tolerant distributed quantum computing with surface codes.

\section{Conclusion and Future Work}\label{conclusion}

In this work, we investigated the logical performance of distributed merge operations under heterogeneous noise by considering a noise model that accounts for errors from both local gates and inter-module entanglement links. Focusing on logical \(Z\) errors during an \(\overline{XX}\) merge between surface-code patches hosted on separate QPUs, we found that the merge threshold remains robust even when the boundary seam is significantly noisier than the bulk. These results show that distributed lattice-surgery operations can tolerate imperfect entanglement links, supporting the feasibility of modular surface-code architectures for fault-tolerant quantum computation.

\vspace{2pt}
Future work can consider detailed, architecture-specific noise models that incorporate realistic physical constraints, including rate-limited entanglement generation, finite quantum memory, and leakage errors. Practical realization of distributed fault-tolerant architectures will require improvements in both hardware models and software protocols, motivating a co-design approach that jointly optimizes physical interconnects, entanglement-generation strategies, and scheduling algorithms. The spatially inhomogeneous noise that arises at distributed seams also motivates the study of decoders beyond standard minimum-weight perfect matching, or weighted variants of it. Comparing belief-propagation methods, union-find variants, and machine-learning-based decoders tailored to biased or spatially varying error profiles is therefore a valuable direction for future work. Finally, while this study focuses on the elementary merge operation, future work should extend the analysis to lattice-surgery based CNOTs, magic state injection, magic-state distillation protocols, and logical teleportation under realistic hardware constraints and diverse network topologies~\cite{pouryousef2026benchmarkingquantumdatacenter}.

\appendix\label{Noisee Model}
We derive effective error rate parameters for a phenomenological noise model tailored to the distributed setting of two QPUs. Specifically, we map underlying physical errors arising from local two-qubit gates, measurements, Bell pairs, and idle times into effective per-cycle error probabilities on the data and measurement (ancilla) qubits.
\vspace{5pt}
We consider the following noise sources:
\begin{itemize}
    \item Local two-qubit gate error: $\varepsilon_{\mathrm{cx}}$
    \item Entanglement pair noise: $\varepsilon_{\mathrm{B}}$
    \item Measurement readout noise: $\varepsilon_{\mathrm{m}}$
    \item Idle noise: $\varepsilon_{\mathrm{idle}}$
\end{itemize}

\vspace{5pt}

We account for the different noise sources contributing to the $Z$ errors for $X$-syndrome extraction. A similar calculation can be done to derive the $X$ errors in the system.
\vspace{5pt}





We model each two-qubit operation (Bell-pair preparation or CNOT gate) as a perfect operation followed by a two-qubit depolarizing channel. This channel applies one of the 15 non-identity two-qubit Pauli operators, each with probability $\varepsilon/15$, where $\varepsilon = \varepsilon_{\mathrm{B}}$ for Bell pairs and $\varepsilon = \varepsilon_{\mathrm{cx}}$ for CNOT gates. The identity operation $II$ is applied with the remaining probability $1-\varepsilon$. We can describe the different error classes associated with this noise channel as:

\vspace{5pt}
\paragraph{Bell pair.}
A Bell state is stabilized by the group $\{II, XX, YY, ZZ\}$ (with the appropriate phases). Two Pauli errors that differ by a stabilizer element act identically on the Bell state. This partitions the 16 two-qubit Pauli operators (including $II$) into four equivalence classes of size four:
\vspace{5pt}
\begin{center}
\begin{tabular}{c c c}
\hline
Representative & Equivalent errors &  Probability \\
\hline
$II$ & $II,\; XX,\; YY,\; ZZ$ & $1 - \tfrac{4}{5}\varepsilon_{\mathrm{B}}$ \\[4pt]
$IX$ & $IX,\; XI,\; ZY,\; YZ$ & $\tfrac{4}{15}\varepsilon_{\mathrm{B}}$ \\[4pt]
$IY$ & $IY,\; XZ,\; ZX,\; YI$ & $\tfrac{4}{15}\varepsilon_{\mathrm{B}}$ \\[4pt]
$IZ$ & $IZ,\; XY,\; ZI,\; YX$ & $\tfrac{4}{15}\varepsilon_{\mathrm{B}}$ \\[4pt]
\hline
\end{tabular}
\end{center}

\vspace{5pt}



\noindent Thus, the 15 non-trivial errors reduce to three effective single-qubit error channels: $IX$, $IY$, and $IZ$, each occurring with probability~$\tfrac{4}{15}\varepsilon_{\mathrm{B}}$. Since $Y \propto XZ$, a $Y$ component on a qubit contributes to both an $X$ and a $Z$ error.
\vspace{5pt}
\paragraph{CNOT gate}

Through Pauli noise propagation, it can be shown that a CNOT gate and a Bell pair share identical error equivalence classes~\cite{Sinclair2025}. To determine the effective error rate on a given qubit, we identify which of the 15 non-identity two-qubit Pauli errors contain an \(X\) or \(Z\) component on that qubit. Assuming a standard depolarizing noise model, where each of these 15 errors occurs with probability~\(\varepsilon_{\mathrm{cx}}/15\), the total probability that a qubit experiences an \(X\) error (or, independently, a \(Z\) error) after a CNOT gate is given by:

\begin{equation}
  p_{X} = p_{Z} = \frac{8}{15}\varepsilon_{\mathrm{cx}}\,.
\end{equation}

\noindent This mapping from the two-qubit depolarizing parameter~$\varepsilon_{\mathrm{cx}}$ (or~$\varepsilon_{\mathrm{B}}$) to effective single-qubit $X$ and $Z$ error rates allows us to parameterize the phenomenological noise model and account for different noise sources. 

\vspace{5pt}

The qubits involved in the merge operation are classified into two distinct regions: 
\begin{itemize}
    \item Bulk region: data and ancilla qubits.
    \item Seam (or boundary) region: data and ancilla qubits.
\end{itemize}
\vspace{5pt}
Let $p_{bulk}$ and $q_{bulk}$ denote the total $Z$-error probabilities per cycle for the data and ancilla qubits in the bulk of the surface code, respectively. Additionally, we define $p_{\mathrm{seam}}$ and $q_{\mathrm{seam}}$ as the corresponding error probabilities for the qubits located along the seam (See circled qubits in Fig.~\ref{fig:merge_operation_QPU}).
\vspace{5pt}


In the absence of additional noise from an external source, errors in the bulk region of the surface code arise from CNOT gate noise~($\varepsilon_{\mathrm{cx}}$), measurements~($\varepsilon_{\mathrm{m}}$), and idle noise~($\varepsilon_{\mathrm{idle}}$). We include idle errors on both data and ancilla qubits to model memory decoherence during inactive time steps in the syndrome-extraction circuit. Each data qubit undergoes four parity-check interactions per code cycle, yielding a cumulative bit-flip or phase-flip probability of,
\begin{equation}
    p_{\mathrm{bulk}} = 4 \left( \frac{8}{15} \right) \varepsilon_{\mathrm{cx}} + \varepsilon_{\mathrm{idle}} \approx{ 3\varepsilon},
\end{equation}

when we consider the uniform noise,
\begin{equation}
\varepsilon_{\mathrm{cx}} =  \varepsilon_{\mathrm{idle}} = \varepsilon.
\end{equation}


Syndrome (ancilla) qubits experience noise similar to that of the bulk data qubits, but additionally have a readout error, leading to,
\begin{equation}
    q_{\mathrm{bulk}} = 4 \left( \frac{8}{15} \right) \varepsilon_{\mathrm{cx}} + \varepsilon_{\mathrm{m}} + \varepsilon_{\mathrm{idle}} \approx 2\varepsilon_{\mathrm{cx}} + \varepsilon_{\mathrm{m}} + \varepsilon_{\mathrm{idle}} \approx 4\varepsilon\,.
\end{equation}




In a rotated surface code, the bulk data qubits (excluding those at the arches or edges) participate in four stabilizer measurements: two $X$-type and two $Z$-type (See Fig.~\ref{fig:surface_code_patch}). Consequently, during syndrome extraction, each data qubit acts as the target for two CNOT gates and as the control for two CNOT gates.


As shown in Fig.~\ref{fig:nonlocal_cnot_decomposition}, $X$ and $Z$-type Pauli errors from the Bell pair propagate asymmetrically through the nonlocal CNOT gate: the target acquires only $X$-type errors, whereas the control acquires only $Z$-type errors. For the circled data qubit (representing the seam data qubits) in Fig.~\ref{fig:merge_operation_QPU}, the two $X$-type stabilizers are shown. However, this qubit is also in the support of the corresponding $Z$-type stabilizers (not shown), which are required for decoding $X$-type errors. This qubit participates in two nonlocal CNOT gates, acting as the control in one and as the target in the other. Consequently, only one of these nonlocal gates contributes additional $Z$-type noise, while the remaining three interactions effectively behave as local gates. This leads to,

\begin{equation}
    p_{\mathrm{seam}} = \underbrace{\left( \frac{8}{15}\varepsilon_{\mathrm{B}} + 2 \left( \frac{8}{15} \right) \varepsilon_{\mathrm{cx}} + \varepsilon_{\mathrm{m}} \right)}_{\text{nonlocal CNOT}} + \underbrace{3 \left( \frac{8}{15} \right) \varepsilon_{\mathrm{cx}}}_{\text{other CNOTs}} + \underbrace{\varepsilon_{\mathrm{idle}}}_{\text{idle }}\,.
\end{equation}
As indicated, the first grouped term represents the noise contribution from the nonlocal CNOT error propagation (resulting from a noisy Bell pair, local CNOTs and a readout error). The second term arises from the remaining three CNOT interactions, and the third term accounts for the idle errors. This expression can be approximated as,


\begin{equation}
    p_{\mathrm{seam}} \approx 0.5\,\varepsilon_{\mathrm{B}} + 2.5\,\varepsilon_{\mathrm{cx}} + \varepsilon_{\mathrm{m}} + \varepsilon_{\mathrm{idle}}\,.
\end{equation}

Assuming a uniform physical error rate~$\varepsilon$ for the local operations (i.e., $\varepsilon_{\mathrm{cx}} = \varepsilon_{\mathrm{m}} = \varepsilon_{\mathrm{idle}} = \varepsilon$) and introducing the Bell-pair noise scaling parameter~$k \geq 1$ such that $\varepsilon_{\mathrm{B}} = k\varepsilon$, this simplifies to,
\begin{equation}
    p_{\mathrm{seam}} \approx \frac{\varepsilon}{2}(k+9)\,.
\end{equation}




\begin{figure}[t]
    \centering
    \includegraphics[width=0.5\columnwidth]{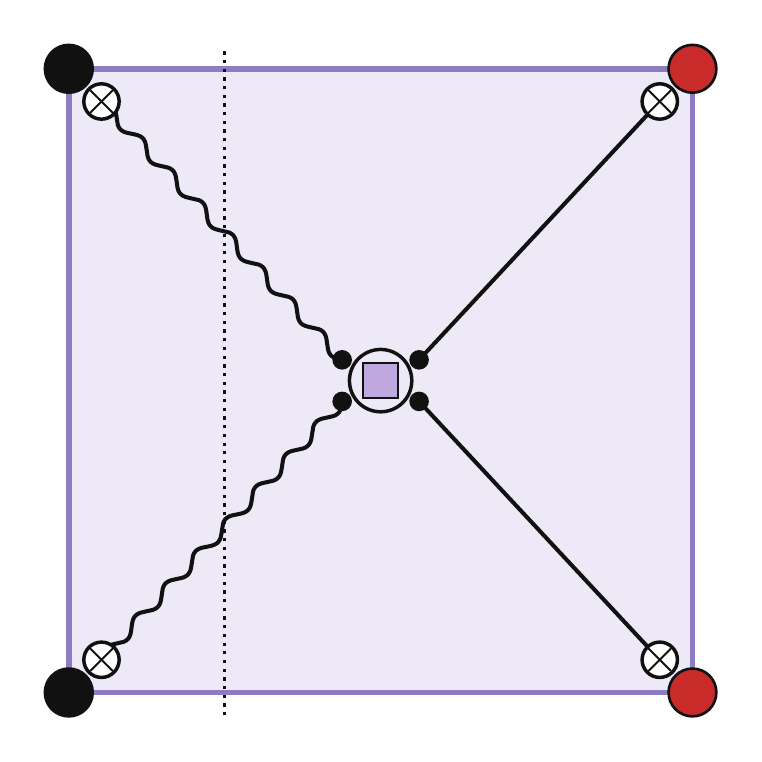}
    \caption{
    Schematic of an $X$-stabilizer in a distributed architecture, where the two left qubits reside in one QPU and the two right qubits reside in another. The corner circles denote data qubits, the central square represents the ancilla qubit, and the connecting lines indicate CNOT interactions. Wavy lines denote nonlocal CNOT gates, while straight lines represent local CNOT gates.
    }
    \label{fig:distributed_stabilizer}
\end{figure}

For the seam ancilla qubits participating in stabilizer measurements, we estimate the effective error rate by identifying the noise sources relevant to $X$-syndrome extraction. An $X$-stabilizer ancilla couples to four surrounding data qubits. Of these four interactions, two are nonlocal; in the corresponding CNOT gates, the ancilla acts as the control qubit (See Fig.~\ref{fig:distributed_stabilizer}). Because $Z$ errors on the control are the relevant error type for $X$-stabilizer measurements, we add all such $Z$-error contributions, yielding,






\begin{equation}
\begin{split}
    q_{\mathrm{seam}} &= \underbrace{2 \left( \frac{8}{15}\varepsilon_{\mathrm{B}} + 2\left(\frac{8}{15}\right)\varepsilon_{\mathrm{cx}} + \varepsilon_{\mathrm{m}} \right)}_{\text{two nonlocal CNOTs}} \\
    &\quad + \underbrace{2 \left( \frac{8}{15} \right) \varepsilon_{\mathrm{cx}}}_{\text{two local CNOTs}} + \underbrace{\varepsilon_{\mathrm{idle}}}_{\text{idle}} + \underbrace{\varepsilon_{\mathrm{m}}}_{\text{readout}}\,.
\end{split}
\end{equation}

We can approximate the above equation as,
\begin{equation}
    q_{\mathrm{seam}} \approx \varepsilon_{\mathrm{B}} + 3\varepsilon_{\mathrm{cx}} + 2\varepsilon_{\mathrm{m}} + \varepsilon_{\mathrm{idle}} + \varepsilon_{\mathrm{m}}\,.
\end{equation}


Finally, for $\varepsilon_{\mathrm{B}} = k\varepsilon$ and a uniform physical error rate for local operations ($\varepsilon_{\mathrm{cx}} = \varepsilon_{\mathrm{m}} = \varepsilon_{\mathrm{idle}} = \varepsilon$), this simplifies to,
\begin{equation}
    q_{\mathrm{seam}} \approx (k+7)\varepsilon\,.
\end{equation}


We note that the error parameters derived above are applied as an approximate noise model, even to some boundary qubits with lower-weight stabilizers. A more refined approach could assign distinct parameters to arches and edges to account for this difference.

\bibliographystyle{IEEEtran}
\bibliography{references}

@article{Preskill2018,
  title = {Quantum Computing in the NISQ era and beyond},
  volume = {2},
  ISSN = {2521-327X},
  url = {http://dx.doi.org/10.22331/q-2018-08-06-79},
  DOI = {10.22331/q-2018-08-06-79},
  journal = {Quantum},
  publisher = {Verein zur Forderung des Open Access Publizierens in den Quantenwissenschaften},
  author = {Preskill,  John},
  year = {2018},
  month = aug,
  pages = {79}
}

@inproceedings{Brandhofer2021,
  title = {Special Session: Noisy Intermediate-Scale Quantum (NISQ) Computers—How They Work,  How They Fail,  How to Test Them?},
  url = {http://dx.doi.org/10.1109/VTS50974.2021.9441047},
  DOI = {10.1109/vts50974.2021.9441047},
  booktitle = {2021 IEEE 39th VLSI Test Symposium (VTS)},
  publisher = {IEEE},
  author = {Brandhofer,  Sebastian and Devitt,  Simon and Wellens,  Thomas and Polian,  Ilia},
  year = {2021},
  month = apr,
  pages = {1–10}
}

@misc{FromNISQtoISQs,
  author = {Juan Miguel Arrazola and Xanadu},
  title = {From NISQ to ISQ},
  year = {2023},
  howpublished = "\url{https://pennylane.ai/blog/2023/06/from-nisq-to-isq/
}",
  note          = {Accessed: 2026-04-14}
}

@misc{pascuzzi2024quantumcentricsupercomputingphysicsresearch,
      title={Quantum-centric Supercomputing for Physics Research}, 
      author={Vincent R. Pascuzzi and Antonio Córcoles},
      year={2024},
      eprint={2408.11741},
      archivePrefix={arXiv},
      primaryClass={quant-ph},
      url={https://arxiv.org/abs/2408.11741}, 
}

@article{Preskill2025,
  title = {Beyond NISQ: The Megaquop Machine},
  volume = {6},
  ISSN = {2643-6817},
  url = {http://dx.doi.org/10.1145/3723153},
  DOI = {10.1145/3723153},
  number = {3},
  journal = {ACM Transactions on Quantum Computing},
  publisher = {Association for Computing Machinery (ACM)},
  author = {Preskill,  John},
  year = {2025},
  month = apr,
  pages = {1–7}
}

@misc{eisert2025mindgapsfraughtroad,
      title={Mind the gaps: The fraught road to quantum advantage}, 
      author={Jens Eisert and John Preskill},
      year={2025},
      eprint={2510.19928},
      archivePrefix={arXiv},
      primaryClass={quant-ph},
      url={https://arxiv.org/abs/2510.19928}, 
}

@article{Chen2023,
  title = {The complexity of NISQ},
  volume = {14},
  ISSN = {2041-1723},
  url = {http://dx.doi.org/10.1038/s41467-023-41217-6},
  DOI = {10.1038/s41467-023-41217-6},
  number = {1},
  journal = {Nature Communications},
  publisher = {Springer Science and Business Media LLC},
  author = {Chen,  Sitan and Cotler,  Jordan and Huang,  Hsin-Yuan and Li,  Jerry},
  year = {2023},
  month = sep 
}

@article{FellousAsiani2021,
  title = {Limitations in Quantum Computing from Resource Constraints},
  volume = {2},
  ISSN = {2691-3399},
  url = {http://dx.doi.org/10.1103/PRXQuantum.2.040335},
  DOI = {10.1103/prxquantum.2.040335},
  number = {4},
  journal = {PRX Quantum},
  publisher = {American Physical Society (APS)},
  author = {Fellous-Asiani,  Marco and Chai,  Jing Hao and Whitney,  Robert S. and Auffèves,  Alexia and Ng,  Hui Khoon},
  year = {2021},
  month = nov 
}

@article{Main2025,
  title = {Distributed quantum computing across an optical network link},
  volume = {638},
  ISSN = {1476-4687},
  url = {http://dx.doi.org/10.1038/s41586-024-08404-x},
  DOI = {10.1038/s41586-024-08404-x},
  number = {8050},
  journal = {Nature},
  publisher = {Springer Science and Business Media LLC},
  author = {Main,  D. and Drmota,  P. and Nadlinger,  D. P. and Ainley,  E. M. and Agrawal,  A. and Nichol,  B. C. and Srinivas,  R. and Araneda,  G. and Lucas,  D. M.},
  year = {2025},
  month = feb,
  pages = {383–388}
}

@book{Cuomo2024,
  title = {Architectures and Circuits for Distributed Quantum Computing},
  ISBN = {9783031738081},
  ISSN = {2190-5061},
  url = {http://dx.doi.org/10.1007/978-3-031-73808-1},
  DOI = {10.1007/978-3-031-73808-1},
  journal = {Springer Theses},
  publisher = {Springer Nature Switzerland},
  author = {Cuomo,  Daniele},
  year = {2024}
}

@article{Kaur2025,
  title = {Optimized Quantum Circuit Partitioning Across Multiple Quantum Processors},
  volume = {6},
  ISSN = {2689-1808},
  url = {http://dx.doi.org/10.1109/TQE.2025.3623158},
  DOI = {10.1109/tqe.2025.3623158},
  journal = {IEEE Transactions on Quantum Engineering},
  publisher = {Institute of Electrical and Electronics Engineers (IEEE)},
  author = {Kaur,  Eneet and Pouryousef,  Shahrooz and Shapourian,  Hassan and Zhao,  Jiapeng and Kilzer,  Michael and Kompella,  Ramana and Nejabati,  Reza},
  year = {2025},
  pages = {1–17}
}

@inproceedings{Chandra2024,
  title = {Network Operations Scheduling for Distributed Quantum Computing},
  url = {http://dx.doi.org/10.1109/TPS-ISA62245.2024.00068},
  DOI = {10.1109/tps-isa62245.2024.00068},
  booktitle = {2024 IEEE 6th International Conference on Trust,  Privacy and Security in Intelligent Systems,  and Applications (TPS-ISA)},
  publisher = {IEEE},
  author = {Chandra,  Nitish K. and Kaur,  Eneet and Seshadreesan,  Kaushik P.},
  year = {2024},
  month = oct,
  pages = {506–515}
}

@article{Monroe2014,
  title = {Large-scale modular quantum-computer architecture with atomic memory and photonic interconnects},
  volume = {89},
  ISSN = {1094-1622},
  url = {http://dx.doi.org/10.1103/PhysRevA.89.022317},
  DOI = {10.1103/physreva.89.022317},
  number = {2},
  journal = {Physical Review A},
  publisher = {American Physical Society (APS)},
  author = {Monroe,  C. and Raussendorf,  R. and Ruthven,  A. and Brown,  K. R. and Maunz,  P. and Duan,  L.-M. and Kim,  J.},
  year = {2014},
  month = feb 
}

@article{Chandra2026,
  title = {Multiplexed Bilayered Realization of Fault-Tolerant Quantum Computation Over Optically Networked Trapped-Ion Modules},
  volume = {7},
  ISSN = {2689-1808},
  url = {http://dx.doi.org/10.1109/TQE.2025.3649617},
  DOI = {10.1109/tqe.2025.3649617},
  journal = {IEEE Transactions on Quantum Engineering},
  publisher = {Institute of Electrical and Electronics Engineers (IEEE)},
  author = {Chandra,  Nitish Kumar and Guha,  Saikat and Seshadreesan,  Kaushik P.},
  year = {2026},
  pages = {1–18}
}

@article{Jacinto2026,
  title = {Network requirements for distributed quantum computation},
  volume = {8},
  ISSN = {2643-1564},
  url = {http://dx.doi.org/10.1103/v9ln-c4v2},
  DOI = {10.1103/v9ln-c4v2},
  number = {1},
  journal = {Physical Review Research},
  publisher = {American Physical Society (APS)},
  author = {Jacinto,  Hugo and Gouzien,  {\'E}lie and Sangouard,  Nicolas},
  year = {2026},
  month = feb 
}

@article{Larasati2025,
  title = {Towards fault-tolerant distributed quantum computation (FT-DQC): Taxonomy,  recent progress,  and challenges},
  volume = {11},
  ISSN = {2405-9595},
  url = {http://dx.doi.org/10.1016/j.icte.2025.03.007},
  DOI = {10.1016/j.icte.2025.03.007},
  number = {3},
  journal = {ICT Express},
  publisher = {Elsevier BV},
  author = {Larasati,  Harashta Tatimma and Choi,  Byung-Soo},
  year = {2025},
  month = jun,
  pages = {417–435}
}

@article{Fowler2012,
  title = {Surface codes: Towards practical large-scale quantum computation},
  volume = {86},
  ISSN = {1094-1622},
  url = {http://dx.doi.org/10.1103/PhysRevA.86.032324},
  DOI = {10.1103/physreva.86.032324},
  number = {3},
  journal = {Physical Review A},
  publisher = {American Physical Society (APS)},
  author = {Fowler,  Austin G. and Mariantoni,  Matteo and Martinis,  John M. and Cleland,  Andrew N.},
  year = {2012},
  month = sep 
}

@article{Nickerson2013,
  title = {Topological quantum computing with a very noisy network and local error rates approaching one percent},
  volume = {4},
  ISSN = {2041-1723},
  url = {http://dx.doi.org/10.1038/ncomms2773},
  DOI = {10.1038/ncomms2773},
  number = {1},
  journal = {Nature Communications},
  publisher = {Springer Science and Business Media LLC},
  author = {Nickerson,  Naomi H. and Li,  Ying and Benjamin,  Simon C.},
  year = {2013},
  month = apr 
}

@article{Terhal2015,
  title = {Quantum error correction for quantum memories},
  volume = {87},
  ISSN = {1539-0756},
  url = {http://dx.doi.org/10.1103/RevModPhys.87.307},
  DOI = {10.1103/revmodphys.87.307},
  number = {2},
  journal = {Reviews of Modern Physics},
  publisher = {American Physical Society (APS)},
  author = {Terhal,  Barbara M.},
  year = {2015},
  month = apr,
  pages = {307–346}
}

@article{Horsman2012,
  title = {Surface code quantum computing by lattice surgery},
  volume = {14},
  ISSN = {1367-2630},
  url = {http://dx.doi.org/10.1088/1367-2630/14/12/123011},
  DOI = {10.1088/1367-2630/14/12/123011},
  number = {12},
  journal = {New Journal of Physics},
  publisher = {IOP Publishing},
  author = {Horsman,  Dominic and Fowler,  Austin G and Devitt,  Simon and Meter,  Rodney Van},
  year = {2012},
  month = dec,
  pages = {123011}
}

@article{Litinski2019,
  title = {A Game of Surface Codes: Large-Scale Quantum Computing with Lattice Surgery},
  volume = {3},
  ISSN = {2521-327X},
  url = {http://dx.doi.org/10.22331/q-2019-03-05-128},
  DOI = {10.22331/q-2019-03-05-128},
  journal = {Quantum},
  publisher = {Verein zur Forderung des Open Access Publizierens in den Quantenwissenschaften},
  author = {Litinski,  Daniel},
  year = {2019},
  month = mar,
  pages = {128}
}

@article{Molavi2025,
  title = {Dependency-Aware Compilation for Surface Code Quantum Architectures},
  volume = {9},
  ISSN = {2475-1421},
  url = {http://dx.doi.org/10.1145/3720416},
  DOI = {10.1145/3720416},
  number = {OOPSLA1},
  journal = {Proceedings of the ACM on Programming Languages},
  publisher = {Association for Computing Machinery (ACM)},
  author = {Molavi,  Abtin and Xu,  Amanda and Tannu,  Swamit and Albarghouthi,  Aws},
  year = {2025},
  month = apr,
  pages = {57–84}
}

@article{Katabarwa2024,
  title = {Early Fault-Tolerant Quantum Computing},
  volume = {5},
  ISSN = {2691-3399},
  url = {http://dx.doi.org/10.1103/PRXQuantum.5.020101},
  DOI = {10.1103/prxquantum.5.020101},
  number = {2},
  journal = {PRX Quantum},
  publisher = {American Physical Society (APS)},
  author = {Katabarwa,  Amara and Gratsea,  Katerina and Caesura,  Athena and Johnson,  Peter D.},
  year = {2024},
  month = jun 
}

@article{m7tq-9v3g,
  title = {Transversal logical Clifford gates on the rotated surface code with reconfigurable neutral atom arrays},
  author = {Chen, Zi-Han and Chen, Ming-Cheng and Lu, Chao-Yang and Pan, Jian-Wei},
  journal = {Phys. Rev. Lett.},
  pages = {--},
  year = {2026},
  month = {Jan},
  publisher = {American Physical Society},
  doi = {10.1103/m7tq-9v3g},
  url = {https://link.aps.org/doi/10.1103/m7tq-9v3g}
}

@article{Vuillot2019,
  title = {Code deformation and lattice surgery are gauge fixing},
  volume = {21},
  ISSN = {1367-2630},
  url = {http://dx.doi.org/10.1088/1367-2630/ab0199},
  DOI = {10.1088/1367-2630/ab0199},
  number = {3},
  journal = {New Journal of Physics},
  publisher = {IOP Publishing},
  author = {Vuillot,  Christophe and Lao,  Lingling and Criger,  Ben and Garc{\'i}a Almud{\'e}ver,  Carmen and Bertels,  Koen and Terhal,  Barbara M},
  year = {2019},
  month = mar,
  pages = {033028}
}

@article{SerraPeralta2026,
  title = {Decoding across Transversal Clifford Gates in the Surface Code},
  volume = {7},
  ISSN = {2691-3399},
  url = {http://dx.doi.org/10.1103/sk5y-25b1},
  DOI = {10.1103/sk5y-25b1},
  number = {1},
  journal = {PRX Quantum},
  publisher = {American Physical Society (APS)},
  author = {Serra-Peralta,  Marc and Shaw,  Mackenzie H. and Terhal,  Barbara M.},
  year = {2026},
  month = feb 
}

@article{Erhard2021,
  title = {Entangling logical qubits with lattice surgery},
  volume = {589},
  ISSN = {1476-4687},
  url = {http://dx.doi.org/10.1038/s41586-020-03079-6},
  DOI = {10.1038/s41586-020-03079-6},
  number = {7841},
  journal = {Nature},
  publisher = {Springer Science and Business Media LLC},
  author = {Erhard,  Alexander and Poulsen Nautrup,  Hendrik and Meth,  Michael and Postler,  Lukas and Stricker,  Roman and Stadler,  Martin and Negnevitsky,  Vlad and Ringbauer,  Martin and Schindler,  Philipp and Briegel,  Hans J. and Blatt,  Rainer and Friis,  Nicolai and Monz,  Thomas},
  year = {2021},
  month = jan,
  pages = {220–224}
}

@article{Lee2022,
  title = {Lattice surgery-based Surface Code architecture using remote logical CNOT operation},
  volume = {21},
  ISSN = {1573-1332},
  url = {http://dx.doi.org/10.1007/s11128-022-03556-z},
  DOI = {10.1007/s11128-022-03556-z},
  number = {6},
  journal = {Quantum Information Processing},
  publisher = {Springer Science and Business Media LLC},
  author = {Lee,  Jonghyun and Kang,  Yujin and Ha,  Jinyoung and Heo,  Jun},
  year = {2022},
  month = jun 
}

@misc{keskin2025latticesurgeryawareresource,
      title={Lattice Surgery Aware Resource Analysis for the Mapping and Scheduling of Quantum Circuits for Scalable Modular Architectures}, 
      author={Batuhan Keskin and Cameron Afradi and Sylvain Lovis and Maurizio Palesi and Pau Escofet and Carmen G. Almudever and Edoardo Charbon},
      year={2025},
      eprint={2511.21885},
      archivePrefix={arXiv},
      primaryClass={quant-ph},
      url={https://arxiv.org/abs/2511.21885}, 
}

@misc{https://doi.org/10.48550/arxiv.2312.01246,
  doi = {10.48550/ARXIV.2312.01246},
  url = {https://arxiv.org/abs/2312.01246},
  author = {Guinn,  Charles and Stein,  Samuel and Tureci,  Esin and Avis,  Guus and Liu,  Chenxu and Krastanov,  Stefan and Houck,  Andrew A. and Li,  Ang},
  title = {Co-Designed Superconducting Architecture for Lattice Surgery of Surface Codes with Quantum Interface Routing Card},
  publisher = {arXiv},
  year = {2023},
  copyright = {arXiv.org perpetual,  non-exclusive license}
}

@misc{https://doi.org/10.48550/arxiv.2603.06513,
  doi = {10.48550/ARXIV.2603.06513},
  url = {https://arxiv.org/abs/2603.06513},
  author = {Liu,  Sitong and Stack,  John and Sun,  Ke and Van Beeumen,  Roel and Monga,  Inder and Klymko,  Katherine and Brown,  Kenneth R. and Saglamyurek,  Erhan},
  title = {Remote Entanglement in Lattice Surgery: To Distill,  or Not to Distill},
  publisher = {arXiv},
  year = {2026},
  copyright = {arXiv.org perpetual,  non-exclusive license}
}

@misc{sunami2025entanglementboostinglowvolumelogical,
      title={Entanglement boosting: Low-volume logical Bell pair preparation for distributed fault-tolerant quantum computation}, 
      author={Shinichi Sunami and Yutaka Hirano and Toshihide Hinokuma and Hayata Yamasaki},
      year={2025},
      eprint={2511.10729},
      archivePrefix={arXiv},
      primaryClass={quant-ph},
      url={https://arxiv.org/abs/2511.10729}, 
}

@article{Mrton2025,
  title = {Lattice surgery-based logical state teleportation via noisy links},
  volume = {7},
  ISSN = {2643-1564},
  url = {http://dx.doi.org/10.1103/ppng-vbqj},
  DOI = {10.1103/ppng-vbqj},
  number = {3},
  journal = {Physical Review Research},
  publisher = {American Physical Society (APS)},
  author = {M{\'a}rton,  {\'A'}ron and Colmenarez,  Luis and B\"{o}deker,  Lukas and M\"{u}ller,  Markus},
  year = {2025},
  month = sep 
}

@article{Domokos2024,
  title = {Characterization of errors in a CNOT between surface code patches},
  volume = {8},
  ISSN = {2521-327X},
  url = {http://dx.doi.org/10.22331/q-2024-12-27-1577},
  DOI = {10.22331/q-2024-12-27-1577},
  journal = {Quantum},
  publisher = {Verein zur Forderung des Open Access Publizierens in den Quantenwissenschaften},
  author = {Domokos,  B{\'a}lint and M{\'a}rton,  {\'A}ron and Asb{\'o}th,  J{\'a}nos K.},
  year = {2024},
  month = dec,
  pages = {1577}
}

@article{Ramette2024,
  title = {Fault-tolerant connection of error-corrected qubits with noisy links},
  volume = {10},
  ISSN = {2056-6387},
  url = {http://dx.doi.org/10.1038/s41534-024-00855-4},
  DOI = {10.1038/s41534-024-00855-4},
  number = {1},
  journal = {npj Quantum Information},
  publisher = {Springer Science and Business Media LLC},
  author = {Ramette,  Joshua and Sinclair,  Josiah and Breuckmann,  Nikolas P. and Vuleti{\'c},  Vladan},
  year = {2024},
  month = jun 
}

@article{Sinclair2025,
  title = {Fault-tolerant optical interconnects for neutral-atom arrays},
  volume = {7},
  ISSN = {2643-1564},
  url = {http://dx.doi.org/10.1103/PhysRevResearch.7.013313},
  DOI = {10.1103/physrevresearch.7.013313},
  number = {1},
  journal = {Physical Review Research},
  publisher = {American Physical Society (APS)},
  author = {Sinclair,  Josiah and Ramette,  Joshua and Grinkemeyer,  Brandon and Bluvstein,  Dolev and Lukin,  Mikhail D. and Vuleti{\'c},  Vladan},
  year = {2025},
  month = mar 
}

@article{Tomita2014,
  title = {Low-distance surface codes under realistic quantum noise},
  volume = {90},
  ISSN = {1094-1622},
  url = {http://dx.doi.org/10.1103/PhysRevA.90.062320},
  DOI = {10.1103/physreva.90.062320},
  number = {6},
  journal = {Physical Review A},
  publisher = {American Physical Society (APS)},
  author = {Tomita,  Yu and Svore,  Krysta M.},
  year = {2014},
  month = dec 
}

@article{Higgott2022,
  title = {PyMatching: A Python Package for Decoding Quantum Codes with Minimum-Weight Perfect Matching},
  volume = {3},
  ISSN = {2643-6817},
  url = {http://dx.doi.org/10.1145/3505637},
  DOI = {10.1145/3505637},
  number = {3},
  journal = {ACM Transactions on Quantum Computing},
  publisher = {Association for Computing Machinery (ACM)},
  author = {Higgott,  Oscar},
  year = {2022},
  month = jun,
  pages = {1–16}
}

@article{Dennis2002,
  title = {Topological quantum memory},
  volume = {43},
  ISSN = {1089-7658},
  url = {http://dx.doi.org/10.1063/1.1499754},
  DOI = {10.1063/1.1499754},
  number = {9},
  journal = {Journal of Mathematical Physics},
  publisher = {AIP Publishing},
  author = {Dennis,  Eric and Kitaev,  Alexei and Landahl,  Andrew and Preskill,  John},
  year = {2002},
  month = sep,
  pages = {4452–4505}
}

@article{Eastin2009,
  title = {Restrictions on Transversal Encoded Quantum Gate Sets},
  volume = {102},
  ISSN = {1079-7114},
  url = {http://dx.doi.org/10.1103/PhysRevLett.102.110502},
  DOI = {10.1103/physrevlett.102.110502},
  number = {11},
  journal = {Physical Review Letters},
  publisher = {American Physical Society (APS)},
  author = {Eastin,  Bryan and Knill,  Emanuel},
  year = {2009},
  month = mar 
}

@misc{kottmann2025latticesurgery,
    title         = {Introducing lattice surgery},
    author        = {Korbinian Kottmann},
    year          = {2025},
    month         = {12},
    journal       = {PennyLane Demos},
    publisher     = {Xanadu},
    howpublished  = {\url{https://pennylane.ai/qml/demos/tutorial_lattice_surgery}},
    note          = {Accessed: 2026-04-14}
}

@article{Wang2003,
  title = {Confinement-Higgs transition in a disordered gauge theory and the accuracy threshold for quantum memory},
  volume = {303},
  ISSN = {0003-4916},
  url = {http://dx.doi.org/10.1016/s0003-4916(02)00019-2},
  DOI = {10.1016/s0003-4916(02)00019-2},
  number = {1},
  journal = {Annals of Physics},
  publisher = {Elsevier BV},
  author = {Wang,  Chenyang and Harrington,  Jim and Preskill,  John},
  year = {2003},
  month = Jan,
  pages = {31–58}
}

@article{ORourke2025,
  title = {Compare the pair: Rotated versus unrotated surface codes at equal logical error rates},
  volume = {7},
  ISSN = {2643-1564},
  url = {http://dx.doi.org/10.1103/PhysRevResearch.7.033074},
  DOI = {10.1103/physrevresearch.7.033074},
  number = {3},
  journal = {Physical Review Research},
  publisher = {American Physical Society (APS)},
  author = {O’Rourke,  Anthony Ryan and Devitt,  Simon},
  year = {2025},
  month = Jul 
}

@article{Higgott2023,
  title = {Improved Decoding of Circuit Noise and Fragile Boundaries of Tailored Surface Codes},
  volume = {13},
  ISSN = {2160-3308},
  url = {http://dx.doi.org/10.1103/PhysRevX.13.031007},
  DOI = {10.1103/physrevx.13.031007},
  number = {3},
  journal = {Physical Review X},
  publisher = {American Physical Society (APS)},
  author = {Higgott,  Oscar and Bohdanowicz,  Thomas C. and Kubica,  Aleksander and Flammia,  Steven T. and Campbell,  Earl T.},
  year = {2023},
  month = Jul 
}

@misc{pouryousef2026benchmarkingquantumdatacenter,
      title={Benchmarking Quantum Data Center Architectures: A Performance and Scalability Perspective}, 
      author={Shahrooz Pouryousef and Eneet Kaur and Hassan Shapourian and Don Towsley and Ramana Kompella and Reza Nejabati},
      year={2026},
      eprint={2601.01353},
      archivePrefix={arXiv},
      primaryClass={quant-ph},
      url={https://arxiv.org/abs/2601.01353}, 
}

@article{Fowler2012proof,
  title = {Proof of Finite Surface Code Threshold for Matching},
  volume = {109},
  ISSN = {1079-7114},
  url = {http://dx.doi.org/10.1103/PhysRevLett.109.180502},
  DOI = {10.1103/physrevlett.109.180502},
  number = {18},
  journal = {Physical Review Letters},
  publisher = {American Physical Society (APS)},
  author = {Fowler,  Austin G.},
  year = {2012},
  month = Nov 
}

@article{Shalby2025,
  title = {Optimized noise-resilient surface code teleportation interfaces},
  volume = {112},
  ISSN = {2469-9934},
  url = {http://dx.doi.org/10.1103/xqrn-wdw1},
  DOI = {10.1103/xqrn-wdw1},
  number = {2},
  journal = {Physical Review A},
  publisher = {American Physical Society (APS)},
  author = {Shalby,  Mohamed A. and Wang,  Renyu and Sedov,  Denis and Pryadko,  Leonid P.},
  year = {2025},
  month = Aug 
}

@inproceedings{Chandra_Color_Code,
  title = {Distributed Realization of Color Codes for Quantum Error Correction},
  url = {http://dx.doi.org/10.1109/QCE65121.2025.00269},
  DOI = {10.1109/qce65121.2025.00269},
  booktitle = {2025 IEEE International Conference on Quantum Computing and Engineering (QCE)},
  publisher = {IEEE},
  author = {Chandra,  Nitish Kumar and Tipper,  David and Nejabati,  Reza and Kaur,  Eneet and Seshadreesan,  Kaushik P.},
  year = {2025},
  month = Aug,
  pages = {2482–2492}
}

\end{document}